# An assessment of estimation models and investment gaps for the deployment of high-speed broadband networks in NUTS3 regions to meet the objectives of the European Gigabit Society

Ferrandis, Jesús. Universidad Politécnica de Madrid, Spain

Ramos, Sergio. Universidad Nacional de Educación a Distancia (UNED), Spain

Feijoo, Claudio. Universidad Politécnica de Madrid, Spain

**Abstract**

This paper analyses the situation in relation to the deployment of high-speed broadband networks in the European Union (EU). Its aim is to assess the investment required to meet the targets set by the European Commission (EC) for 2025, within the framework of the European Gigabit Society (EGS). This plan aims to ensure the availability and take-up of very high-capacity fixed and wireless networks, in both urban and rural areas, among households and the main socio-economic drivers. The estimation model presented here uses a methodology supported by data at the local (NUTS3) level to give a bottom-up estimation of the investment gap for each of the EGS objectives, using three different scenarios depending on the mix of wired and wireless technologies offered. The methodology and estimation model used in the paper are examined against other examples and assumptions available in the literature. We also offer a dynamic perspective on the analysis of the evolution of this investment gap over the years 2017 to 2019, which includes an assessment of the usefulness of these estimation models. The paper concludes by identifying the need for new measures to attract investment from public and private sources, mostly in rural areas, since purely market-driven initiatives fall short by about two-thirds of the required total investment. We also highlight how estimation models, despite their apparent complexity and number of assumptions, have a proven capacity to provide sufficiently accurate figures to guide policy action at the local, regional, national and supra-national levels. The usefulness of these estimation models could be enhanced by the availability of homogeneous and stable information on socio-demographics and on the level of coverage and take-up of broadband networks, at least at NUTS3 level.

**Keywords**

High-speed broadband networks, European Gigabit Society, fibre networks, wireless networks, 5G, investment gap, rural areas, broadband networks deployment models networks that are able to provide high-speed Internet connectivity (see Abrardi and Cambini (2019) for an extensive review of the literature, with a particular focus on the results and methodologies of the most recent studies).

In fact, many governments around the world have already decided to directly intervene in the deployment of broadband networks, in a search for contributions to the economy and for social enhancement. They seem particularly concerned with inequities in terms of access to telecommunications infrastructure and the lack of opportunities that its absence entails, as shown in the cases of Australia, New Zealand (Given, 2010), Korea (Hong, 2017; Gómez-Barroso & Feijóo, 2009) and the UK (Galloway, 2007) (see also Ruhle et al., (2011) for an international comparison).

## 1.1. Digital Agenda for Europe

In the European Union (EU), based on the same rationale of increasing social and economic prosperity, the European Commission (EC) started monitoring the evolution of broadband connectivity across EU Member States, and subsequently established initiatives to improve it, with a focus on areas that lacked communications network infrastructures. In the first stage, the Digital Agenda for Europe (DAE) was established in May 2010, which included a set of targets for the availability of basic and high-speed broadband connectivity in European households by 2020. The two goals of DAE were "to ensure that, by 2020, (i) all Europeans will have access to internet speeds of above 30 Mbps, and (ii) 50% or more of European households will subscribe to internet connections above 100 Mbps" (European Commission, 2010). This initiative was launched at a time when Europe was noticeably lagging behind the leading Asian countries in terms of the coverage and adoption of ultrabroadband fibre-based networks (Briglauer, 2014).

Around this time, it also became evident that the amount of investment required to fulfil these goals exceeded the market players' willingness to invest, and that some regulatory push, if not direct public support, was required (Gómez-Barroso & Feijóo, 2012). According to estimations made by the EC (Analysis Mason & Tech4I2, 2013), achieving the first goal of 100% coverage at 30 Mbps would require an investment of between 27 b€ and 250 b€, depending on the technology chosen[1]. Similarly, Hätönen (2011) followed a granular, bottom-up approach in which the cheapest technology was chosen for deployment in every area[2], and estimated that meeting the targets for full coverage in the EU would require between 73 and 221 b€, based on four different scenarios depending on speed requirements (theoretical vs. actual, asymmetric vs. symmetric) and access requirements (communal vs. households). Along the same lines, Feijóo et al. (2018) also concluded that at that time, "the EU in general was far from achieving the DAE targets in 2020 in terms of infrastructure deployment […] putting the figure on the investments needed to fill the gap at around 50 b€ for the EU-28 beyond already expected public and private funding, with more than 90% of the pending investments in rural geotypes".

---

[1] Cost of deploying fixed networks based on fibre to cover 100% households would be 45 b€ in the case of fibre to the cabinet (FTTC) and 250 b€ in the case of fibre to the premises (FTTP). If wireless instead of fixed technologies were deployed, the cost would be 42 b€ for HSPA+ and 27 b€ for LTE.

[2] Areas for this study are equivalent to NUTS4 regions according to the EU's Nomenclature of Territorial Units for Statistics (NUTS).

During the period in which the DAE's objectives guided high-speed broadband policy action in EU Member States, reports from market analysts and the academic literature identified a number of pending issues, in particular: (i) the need to quantify in adequate detail and monitor the amount of investment required to fill the broadband investment gap, as the EC launched calls for tenders in this area (Analysis Mason & Tech4I2, 2013); (ii) the exact areas in which this investment was needed due to a lack of market interest, i.e. a "white, grey and black areas analysis" (Feijóo, Gómez-Barroso, & Bohlin, 2011); and (iii) the most suitable strategies for addressing the broadband investment gap, adapted to specific countries and specific regions within a country, to create positive broadband market dynamics that were notoriously "path-dependent" (Lemstra & Melody, 2015). In short, problems remained as to how much investment, where and how.

As a consequence, these three intertwined issues attracted considerable interest from academicians and institutions, giving rise to a range of methodologies, results and publications (see Feijóo et al. (2018) for a summary of issues related to DAE). An overarching examination of these contributions revealed two major conclusions: (i) the need for an additional public push within a framework of very high levels of investment to complete the deployment of high-speed broadband networks in the EU; and (ii) the need for the availability of relevant data, robust network deployment models and a continuous monitoring exercise to maintain an up-to-date picture of the pending deployments and investment gap.

## 1.2. European Gigabit Society

Against this background, new technologies were developed that rendered obsolete some of the deployment scenarios considered by the DAE, despite the fact that the period allocated for their completion had not yet ended. Thus, the European Commission adopted a new set of initiatives and legislative proposals on September 2016 that would place the EU at the forefront of Internet high-speed connectivity. In particular, the strategy on Connectivity for a European Gigabit Society (EGS) (European Commission, 2016) presented a vision for Europe in which the availability and take-up of very high-capacity telecommunications networks enabled the widespread use of products, services and applications with the Digital Single Market, and the goal of incentivising investment in new high-speed broadband infrastructure (Briglauer, Stocker, & Whalley, 2020).

More specifically, the EGS initiative proposed an interim target for 2020, as follows:

- Intermediate 5G connectivity to be available as a fully-fledged commercial service in at least one major city in each EU Member State (Target 1, T1).

It also proposed three main strategic objectives for 2025:

- 5G connectivity: All urban areas[3] and all major terrestrial transport paths[4] to enjoy uninterrupted 5G coverage (Target 2, T2);
- Gigabit connectivity: Gigabit connectivity for all main socio-economic drivers such as transport hubs, and main providers of public services[5] as well as digitally intensive enterprises[6] (Target 3, T3);
- Rural connectivity: All European households, rural or urban, to have access to Internet connectivity offering a downlink of at least 100 Mbps, upgradable to gigabit speed (Target 4, T4).

A relevant novelty was that for the first time, EGS mentioned a particular technology, in relation to the deployment of 5G in Targets 1 and 2 above, in stark contrast with DAE, although still a part of EGS remained technology-agnostic, as is shown in Targets 3 and 4 above. This could be seen as an attempt by the EC to help boost the adoption of 5G in Europe, which is currently falling behind other regions, particularly in Asia[7]; only during 2020 has it shown the first signs of recovery, with the launch of commercial services in 18 countries and the publication of national 5G roadmaps by 12 Member States and the UK[8].

Within this context, this paper aims to provide a bottom-up estimation of the investment gap for the four EGS targets individually and in combination, based on scenarios depending on different mixes of wired and wireless technologies, and in two different moments in time to determine its evolution and assess the validity and usefulness of the estimation exercise. As in the case of the DAE, this type of high-speed broadband investment estimation, and the underlying models, are expected to help in identifying the challenges addressed by policies intended to meet EGS targets.

With the above goal, the paper is structured as follows. After giving some background on EU high-speed broadband initiatives, the next section briefly reviews the existing literature on estimation models for the broadband investment gap. The subsequent section gives an overview of the deployment status of fixed and mobile broadband networks in the UE, which forms the point of departure for any estimation model of the investment gap. Next, the paper discusses the methodology and assumptions used to estimate the EGS gaps for 2025 and makes explicit the strategic choices that guide the analysis. Next section describes the resulting investment gaps for each individual target and their combinations, in different technological scenarios; a discussion and a comparison with other studies are also presented. The paper closes with the main conclusions of the study, with a focus on the expected development of high-speed broadband networks in 2025 and the need for new approaches and further public support.

---

[3] See: https://ec.europa.eu/eurostat/statistics-explained/index.php/Archive:European_cities_–_the_EU-OECD_functional_urban_area_definition

[4] Motorways, national roads and railways, in line with the definition of Trans-European Transport Networks.

[5] This includes primary and secondary schools, train stations, ports and airports, local authority buildings, universities, research centres, doctors' surgeries, hospitals, stadiums, law-related facilities, etc.

[6] Enterprises with a high level of integration of digital technologies.

[7] The first commercial launch of 5G took place in Korea in April 2019. By May 2020, 73 mobile operators in a total of 41 countries had launched 5G services (see Analysys Mason's report "73 MNOs worldwide have launched 5G services, bringing benefits to consumers and communities"). In Europe, the first such launch was Sunrise, the second-largest player in Switzerland, which focused on offering a fibre-like service in rural and suburban areas, where a large number of customers still use slower DSL technologies. The aim was to signal new hope that 5G would offer an alternative to fixed technologies in places where they were not technically feasible or economically viable. In addition, by May 2020, the portfolio of commercially available 5G devices included 77 phones, 16 CPEs and five laptops, among others (https://gsacom.com/paper/5g-devices-ecosystem-member-report-june-2020/).

[8] According to the 5G Observatory, during 2020 this 12 Member States are Austria, Czechia, Denmark, Estonia, Finland, France, Germany, Luxembourg, Portugal, Spain, Sweden and The Netherlands.

## 2. Literature review of estimation models for high-speed broadband investments

As in the case of the DAE, academics and various institutions have started to analyse the EGS framework to determine possible paths towards meeting its objectives and the associated costs. Although some of these contributors have estimated some of the costs related to network deployment, from the present authors' perspective these only partially cover the EGS objectives and/or focus only on certain options in terms of technology. In addition, as discussed above, the methodologies and scenarios for the deployment of high-speed broadband should form part of a continuous monitoring exercise, in view of the rapid changes in both technologies and policies. This is even more important when the case of 5G is included, due to the uncertainty around the exact form that this technology will take[9]. In the following paragraphs, a succinct summary of previous attempts at addressing the EGS-related investment gaps is presented.

Fibre deployments have attracted most of the initial interest in this area. For example, a report by the consulting firm BCG for ETNO (Bock & Wilms, 2016) estimates that an investment of 307 b€[10] would be required in order to deploy fibre to all of the households in the EU as a way to ensure they have access to ultrafast Internet connectivity[11]. However, this report lacks a detailed description of the model used, and considers neither the cost of reaching enterprises and other institutions nor the possibility of combining the deployment of fibre with the use of wireless technologies, which may be an option in some scenarios. In relation to the deployment of 5G infrastructure, it estimates an investment of 200 b€, but it is not clear how this deployment matches any of the EGS objectives.

Along the same lines, the FTTH Council Europe (2017) has estimated the additional investment required to reach 100% of EU-28 households with fibre[12] and to connect 50% of them. It applied a granular approach using data on population density at the NUTS3 level, including corrections based on issues specific to each country, such as labour rates, and excluding unpopulated areas from the analysis. As a result, it estimated a required additional investment of 156 b€, which could be reduced to 137 b€ through the reuse of existing ducts and in-building infratructure, and through the coordination and sharing of civil works. However, the impact of the potential use of wireless technologies and the costs of reaching enterprises and other institutions are not given in the report.

---

[9] Hypotheses related to the deployment of 5G are presented in a later section. In a bottom-up model, hypotheses can easily be modified to reflect the exact features of the technology that is finally deployed.

[10] This total figure is calculated based on the hypothesis that 53 b€ will be required to upgrade 81 million households from existing copper-based very high-speed digital subscriber line (VDSL) to FTTP, with costs ranging from 720 € to 1,980 € per household depending on geographical density. The remaining sum of 307 b€ is dedicated to the deployment of new FTTP in a further 135 million households, at a cost of between 720 € and 3,150 € per household.

[11] Equivalent to Target 4 in our paper, as discussed in later sections.

[12] Equivalent to Target 4 in our paper, as discussed in later sections.

As part of the documentation for the EGS initiative, a study carried out by Analysys Mason for the EC (Analysys Mason, 2016) examined the EGS objectives in a wider context. It defined a path towards wired and wireless network deployment based on six different scenarios that would cover all of the EGS objectives, both for households and for other socio-economic drivers and professionals (SEDPs)[13]. Considering the combination of all the targets, equivalent to meet all the EGS objectives, this would cost a total of 502 b€. The study also estimated that commercially viable deployment from operators would cover 25% of this investment, leaving a considerable gap that might require alternative funding mechanisms.

A relevant sub-theme within investment estimation models of fibre deployment relates to the effects of competition and the impacts of diverse regulatory actions. For instance, Queder (2020b) uses a discounted cash flow model to compare the financial attractiveness of wholesale-only and integrated business models for a greenfield FTTP rollout. The effects on deployment of the competition between fibre and cable have also been investegated (Queder, 2020a), as well as potential new regulatory regimes such as network neutrality (Charalampopoulos, Katsianis, & Varoutas, 2020) and the impacts of local effects and demand uncertainty on the rollout of high-speed broadband networks (Sahebali et al., 2019). Estimation models can be also used in reverse by regulators to calculate the cost of an efficient operator investing in high-speed broadband networks and to determine appropriate regulations (Logothetis et al., 2019).

A second area of interest for academics and practitioners within this research domain is the deployment of high-speed broadband in rural areas. This area of research is related to the public good features of broadband deployments (Cambini & Jiang, 2009) and to public-private partnerships in telecom infrastructures (Gomez-Barroso & Feijóo, 2010), and specifically accounts for and evaluates the impact of ultra-fast broadband investment or penetration on various economic dimensions and the role of alternative modes of regulation (Abrardi & Cambini, 2019b). In particular, researchers have been attracted to analyses of the investment gap in comparison with urban areas (Lucendo-Monedero, Ruiz-Rodríguez, & González-Relaño, 2019), to the assessment of different policy measures that could contribute to reducing this type of rural-urban digital divide, such as community-based deployments (Salemink & Strijker, 2018), and to the effects of supply-side policies in particular geographies (Hernández, Aguilar Chinea, & Baquero Pérez, 2020). Within this field, an important paper by Rendon, Schneir and Xiong (2016) analyses potential scenarios for rolling out fixed broadband networks to cover 9.3 million homes in rural areas, and develops a model to estimate the costs involved[14], arriving at a total of 33.6 b€ to pass all homes considered with fibre to the home (FTTH) and to connect 50% of them, which would mean a cost of between 2,400 and 3,600 € per home connected.

---

[13] The study includes an estimation of the costs for each individual scenario and for the aggregation of the six scenarios, taking into account potential synergies among them to avoid the double counting of certain costs. Providing mobile connectivity for 95% of the population (at a certain extent equivalent to Target 1 and a proportion of Target 2 in our paper), through a combination of macro mobile network and small cells, would require 175 b€. Adding wireless connectivity to all transport links (equivalent to the rest of Target 2 in our paper) would cost 103 b€, assuming that 75% of railways and motorways, and 50% of other roads are already covered. Deployment of wireline gigabit networks would require 64 b€ to provide connectivity to large SEDPs (including universities, schools, research centres, business parks, innovation hubs, hospitals, libraries, and SMEs employing at least 50 workers), an additional 197 b€ to add small SEDPs (including other SMEs and professionals/freelancers/teleworkers, equivalent to Target 3 in our paper) and another 249 b€ to provide connectivity to residential premises (equivalent to Target 4 in our paper). A comparison with this study will be presented in a later section.

Next, given the high cost of deploying fixed technologies in rural areas, many studies over the last twenty years (starting with the arrival of 3G and more frequently following the emergence of 4G/LTE) have analysed the ability of wireless technologies to serve as enabling, complementary or substitute technologies for fixed broadband services. For instance, a detailed analysis by Ovando, Pérez, & Moral (2015) evaluates the feasibility of using long term evolution (LTE) to provide fixed broadband services in rural areas of Spain, and derives a required capital expenditure ranging from just 860 € to 1,045 € per home, including equipment on the customer's premises, which could be reduced by a further 10% by the sharing of passive network elements. Fixed wireless access (FWA) networks based on LTE technology have also been examined for use as a "last mile" solution to provide high-speed broadband access to areas where fixed broadband is limited (Ioannou, Katsianis, & Varoutas, 2020). In theory, LTE technology offers high-speed connections of up to 300 Mbps, depending on network load and sharing. While this is not sufficient to meet the EGS targets, it can be considered a viable intermediate step as an alternative to other fixed network solutions, especially when considering future upgrades to 5G networks that promise gigabit speeds per user, particularly in rural areas. In a similar vein, Chiha et al. (2020) examine the viability of using a combination of 4G and satellite links as the core of the network. In their model, the average cost per user ranges between 12 to 155 €, depending on the bitrate (level of quality) per user.

Based on these results, we apply a granular methodology starting at the local NUTS3 level and with information from the municipal NUTS5 level, and with a special focus on the particularities of rural geotypes, in order to estimate the existing gaps in connectivity, evaluate alternative options in terms of updated technologies (fixed vs. wireless) and assess their market situation and prospects. We also calculate the investment gaps for 2018 (based on 2017 deployment data) and for 2020 (based on 2019 deployment data), their evolution with respect to the 2025 gap, and the potential impact of different technological options as of 2020. In addition to these updated, more precise calculations and a consideration of possible technology deployment scenarios, our calculations at two different points in time allow to assess the validity and usefulness of these bottom-up estimation models for investment gaps at NUTS3 level.

---

[14] This considers different technological options based on new deployments of fibre, and in certain cases the reuse or existing copper lines (fibre to the home (FTTH), fibre to the distribution point – Building (FTTdp-Building), fibre to the distribution point – Street (FTTdp-Street), fibre to the remote node (FTTRN), fiber to the cabinet (FTTC) and central office – very high-speed digital subscriber line (CO-VDSL)), with six different geotypes depending on the density of premises, the distribution of homes/flats and the distance to the central office.

## 3. Overview of high-speed broadband deployment in the EU

The point of departure for all estimation models and the ensuing analysis should be the status of high-speed broadband deployment in the EU, and in particular those deployments that already meet EGS targets and/or that can be reused to extend the reach of new deployments as required. In fact, the EGS specifically mentions that the existing network infrastructure "needs to be extended by a smart mix of wireless and wireline technologies requiring large investments in broadband infrastructure" (European Commission, 2016). For these reasons, we present an assessment of the status of high-speed broadband deployment in the EU in the following paragraphs.

As part of the key goal of EGS, 32% of EU households were covered by networks providing at least 1 Gbps by mid-2019, with Malta having already achieved 100% coverage; however, in some other countries such as the UK, Romania and Latvia, coverage was at 0% (see Fig. 1)[15].

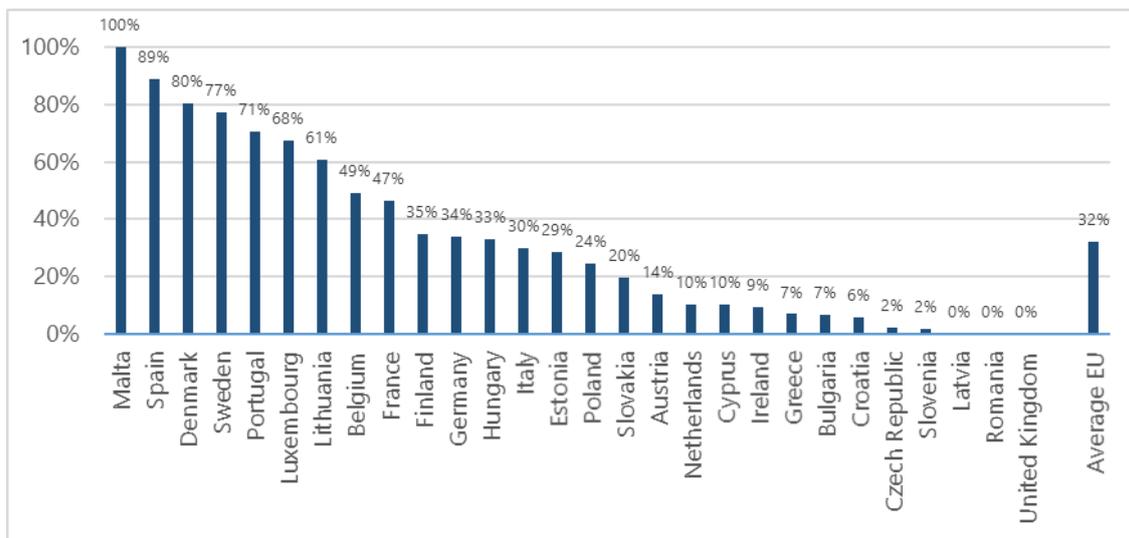

**Fig. 1.** Coverage of broadband networks with at least 1 Gbps in the EU (2019). Source: Adapted from IHS Markit, Point Topic, & Omdia (2020).

While the goal remains connectivity of at least 1 Gbps, a potentially valid starting point for the reuse of high-speed broadband deployments is the existence of wired networks capable of providing downlink speeds of 100 Mbps, which already cover 68% of EU households, 20 percentage points more than in 2015. Of the EU Member States, there are six countries with coverage of above 90%: Malta, Belgium, the Netherlands, Luxembourg, Denmark and Latvia. In contrast, coverage in Bulgaria, Greece and Croatia was below the 50% mark (see Fig. 2).

---

[15] Data used in this chapter are the latest available for a proper comparison at the time of writing.

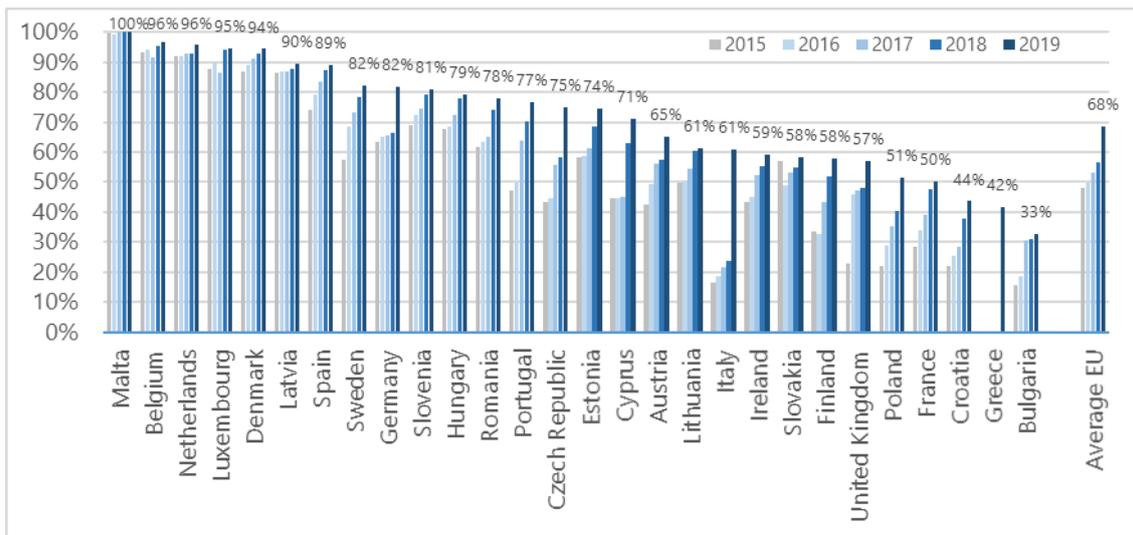

**Fig. 2.** Coverage of broadband networks with at least 100 Mbps in the EU (2019). Source: Adapted from IHS Markit, Point Topic, & Omdia (2020).

Networks that are of particular interest are those that can be easily be upgraded into the Gbps range, i.e. networks based on FTTP and DOCSIS[16] 3.x. In 2019, 34% of households in the EU were ready for upgrade to FTTx with 1 Gbps. Coverage of FTTP exceeded 80% only in two countries, Latvia and Spain, and was below the 10% mark in Belgium and Greece (see Fig. 3). In this regard, it should be noted that this paper follows Ioannou, Katsianis, & Varoutas (2020), among others, in considering that the new EGS target of 100 Mbps connections that are upgradable to 1 Gbps for all households by 2025 is unlikely to be achieved by VDSL vectoring and other copper-based technologies.

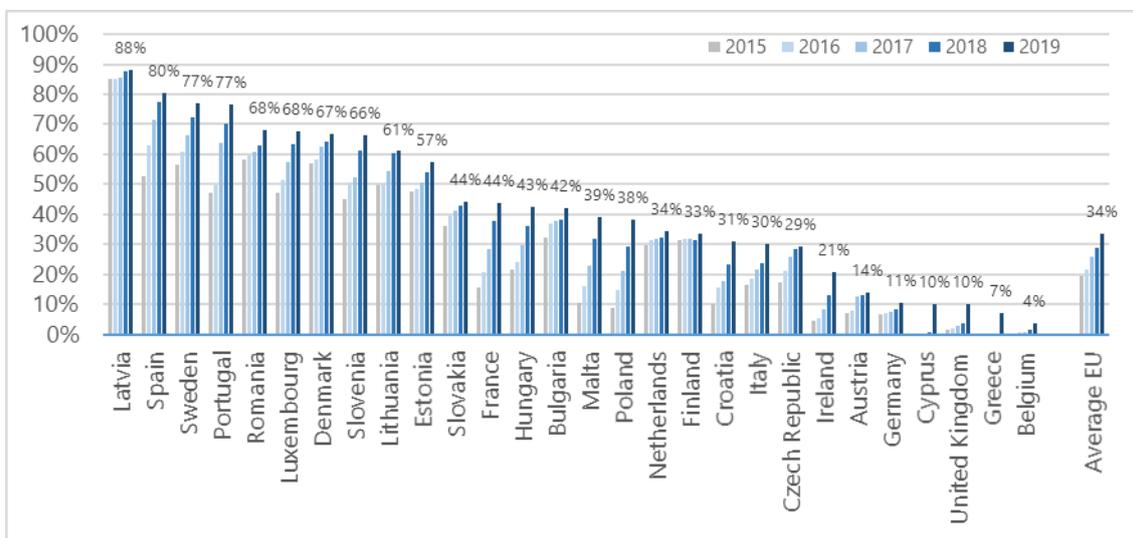

**Fig. 3.** Coverage of FTTP in EU Member States in 2019. Source: Adapted from IHS Markit, Point Topic, & Omdia (2020).

For DOCSIS 3.0, the average coverage in EU was 46% in 2019, a value that has remained stable over several years, highlighting that greenfield deployments no longer use this technology. The figures for coverage in Member States are rather disparate: for Malta, Netherlands and Belgium this is above 90%, while Italy and Greece have negligible coverage (see Fig. 4).

---

[16] Data Over Cable Service Interface Specifications.

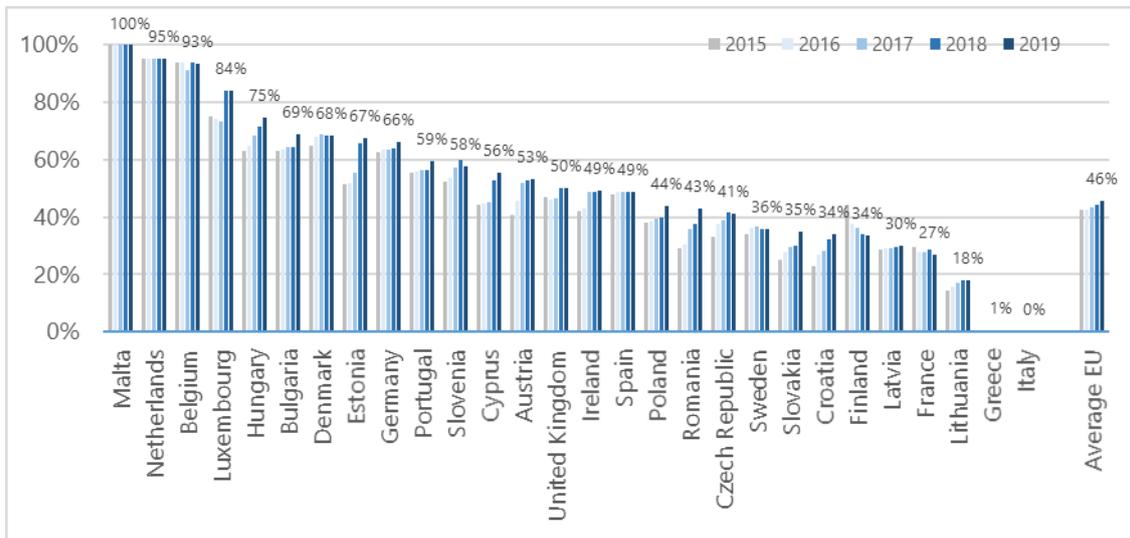

**Fig. 4.** Coverage of FTTP in EU Member States in 2019. Source: Adapted from IHS Markit, Point Topic, & Omdia (2020).

As of 2020, the path to 5G is becoming standardised around a technology called 5G New Radio (5G NR), also known as Release 15, which will eventually be replaced by Release 16. Hence, 5G is an evolving technology with standard equipment for both the network and the user, which has only been available since early 2019. With respect to the objectives related to the deployment of 5G wireless networks (intermediate connectivity (T1) and 5G connectivity (T2)), a relevant point of departure is the current status of deployment of 4G networks, since in most estimation models their footprint will be reused in the upgrade to 5G. In mid-2019, the EU coverage of LTE (4G) was 99%, with all EU Member States having achieved almost full coverage of the population (see Fig. 5).

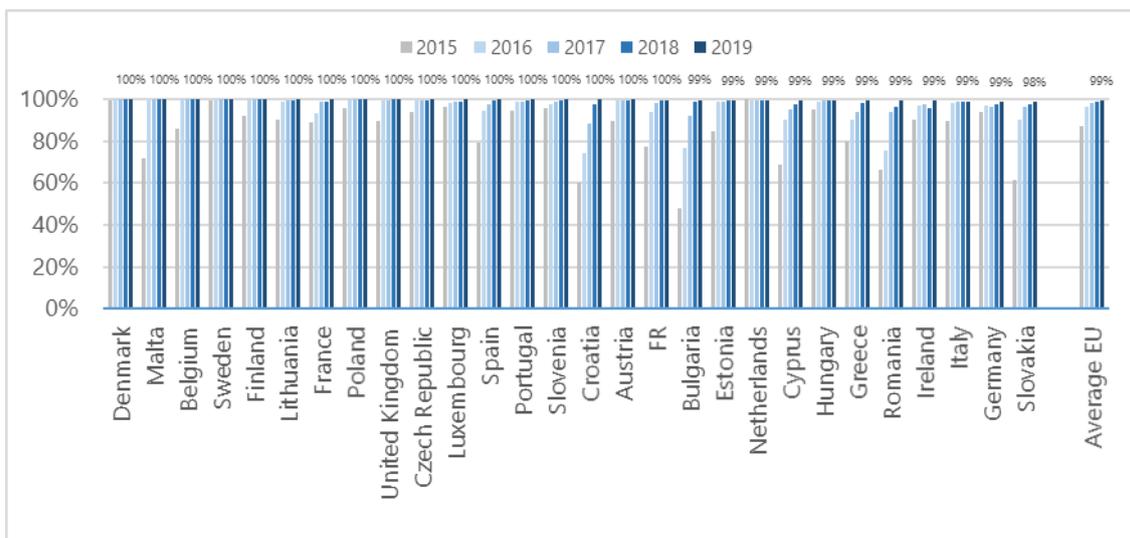

**Fig. 5.** Coverage of LTE in EU Member States in 2019. Source: Adapted from IHS Markit, Point Topic, & Omdia (2020).

In addition to these substantial differences between EU Member States in terms of network deployment, there is also a considerable gap between urban and rural areas[17] across the EU, in relation to both coverage and take-up. According to a report by the European Court of Auditors (2018), the high-speed broadband gap was especially significant in rural areas, where coverage was only 47% of households in 2016. Only three countries (Malta, Luxembourg and the Netherlands, i.e. countries that are small in size or highly urbanised) had rates of rural and urban coverage that were relatively equivalent, and in 14 countries across the EU, the rural coverage was below the 50% threshold. In 2019, the gap in the coverage of technologies that could provide 100 Mbps was 38% in urban areas vs. 18% in rural areas for FTTP, and 22% vs. 4% for DOCSIS 3.1. Other studies have estimated the digital broadband divide between regions at 37%, including both coverage and usage (Lucendo-Monedero, Ruiz-Rodríguez, & González-Relaño, 2019). However, this gap was almost insignificant for LTE (100% urban vs. 98% rural areas), highlighting the convenience of wireless network deployments in rural areas (see Fig. 6).

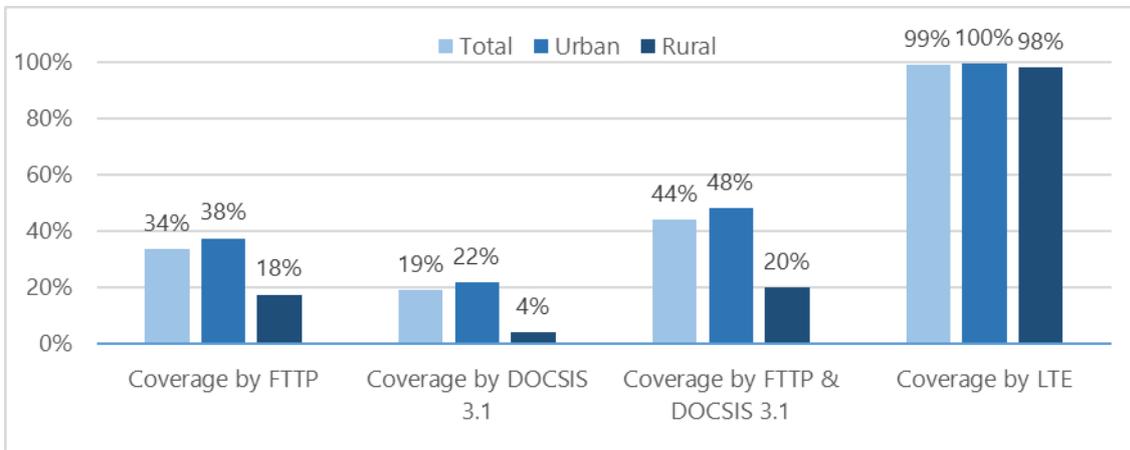

**Fig. 6.**   Coverage in the EU of broadband networks able to provide at least 100 Mbps to households in urban and rural areas, in 2019. Source: Adapted from IHS Markit, Point Topic, & Omdia (2020).

The last element in our overview of high-speed broadband deployment involves the ongoing initiatives and future commitments of operators and public institutions in the EU Member States regarding 1 Gbps broadband networks. A compilation of these is given in Appendix A at the end of this paper.

In summary, most of the countries in the EU seem to be evolving at a slower pace than required to meet the EGS goals, thus highlighting the need for detailed (local, regional, country-wide) estimations of the high-speed broadband investment gap for a range of targets and scenarios.

## 4.  Methodology and estimation model

In this section, the methodology and estimation model used in this paper is explained in some detail. The objective is to make explicit and to discuss all the assumptions made in our estimation of high-speed broadband network investments.

---

[17] See the definitions of urban and rural areas given in the methodology section.

The methodology described below is similar to that followed in a previous work that aimed to estimate the investment required to meet the DAE targets (Feijóo et al., 2018). However, in this paper, the original methodology is updated and reformulated to include new variables and specific issues introduced by the requirements of the EGS targets, and to highlight and discuss the validity and trade-offs made in relation to each of the assumptions used.

Any methodology for the estimation of a high-speed broadband network can be divided into three main stages: (i) establishing the precise status of existing broadband deployments, in order to acknowledge, re-use or discard them; (ii) defining a technical path to meet the EGS objectives; and (iii) estimating the EGS investment gap itself (see Fig. 7).

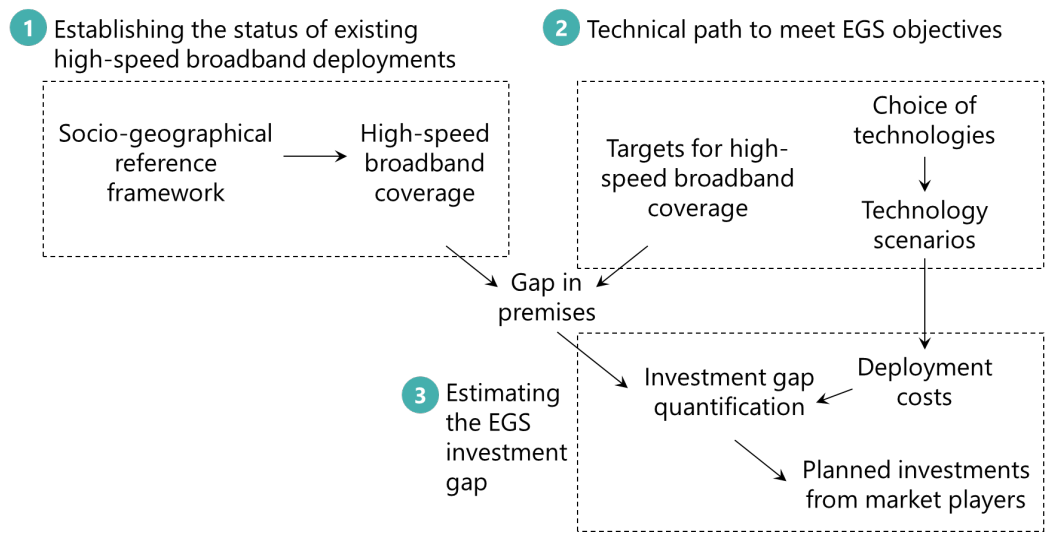

**Fig. 7.** Schematic diagram of the methodology used in this article

## 4.1. Establishing the status of existing high-speed broadband deployments

The starting point for any estimation model is based on (i) the definition of the socio-geographical framework of reference, and (ii) the gathering of existing data of fixed and wireless high-speed broadband coverage within this socio-geographical reference framework.

### 4.1.1. Socio-geographical reference framework

In this paper, we argue that it is possible to use information for the NUTS3 region in the EU-28 as the reference framework. This level of granularity is the result of a trade-off between the complexity of the model and the availability of homogeneous information across the EU-28.

NUTS3 information on population, land area and numbers of households and enterprises is available from the Eurostat database[18]. Due to a lack of more granular data, we also assume in this paper that the households and companies within each country follow the same distribution as the population across NUTS3 areas. This hypothesis was also used by Analysys Mason (2016), and implies that more populated areas also contain a higher concentration of enterprises.

However, it is possible to "open" each NUTS3 region and to characterise it as a set of smaller, municipal-level LAU2 or NUTS5 areas of five different geotypes (urban, suburban, semi-rural, rural and extremely rural). This is made possible through the use of data on the degree of urbanisation (DEGURBA), area and population[19]. An overview of the resulting socio-geographical framework for the whole EU is given in Tables 1 and 2.

**Table 1**

Distribution of NUTS3 regions in terms of population density and geotype (2019)

| Pop. density | Number of NUTS3 regions | Pop. (m) | Number of premises (m) | Urban geotype (%) | Suburban geotype (%) | Semi-rural geotype (%) | Rural geotype (%) | Extremely rural geotype (%) |
|---|---|---|---|---|---|---|---|---|
| Above 500 inh/km2 | 295 | 168.36 | 78.37 | 78.0% | 19.3% | 2.1% | 0.5% | 0.1% |
| 100 to 500 inh/km2 | 559 | 224.48 | 111.03 | 28.1% | 46.9% | 21.3% | 3.6% | 0.1% |
| 50 to 100 inh/km2 | 293 | 92.30 | 43.53 | 22.1% | 32.1% | 25.8% | 19.2% | 0.7% |
| 10 to 50 inh/km2 | 184 | 36.33 | 18.01 | 16.9% | 33.8% | 11.1% | 32.0% | 6.2% |
| Up to 10 inh/km2 | 17 | 2.56 | 1.39 | 12.9% | 36.5% | 4.3% | 14.2% | 32.2% |
| Total | 1,348 | 524.03 | 252.33 | 42.3% | 34.5% | 15.1% | 7.4% | 0.8% |

*Source: Own calculations from Eurostat data*

Table 1 shows that although there is an obvious correlation between the population density of each NUTS3 area and the distribution of geotypes, an assumption that each NUTS3 area belongs only to one particular geotype (as in simpler estimation models) overlooks many interesting situations. For instance, in the more densely populated NUTS3 regions, rural areas make up a significant part of the premises (25% in the case of regions with 100 to 500 inh/km$^2$), while in less densely populated NUTS3 regions, more than 50% of the premises belong to the urban or suburban geotypes.

---

[18] http://appsso.eurostat.ec.europa.eu/nui/show.do?dataset=demo_r_d3area&lang=en (latest update of information for the year 2016)

[19] https://ec.europa.eu/eurostat/web/degree-of-urbanisation/background. Note that DEGURBA classification divides LAU2 regions into only urban, suburban and rural; to increase the granularity of the analysis, we go a step further and introduce a subclassification for rural LAU2 regions, dividing them into semirural, rural and extremely rural depending on their density of population.

**Table 2**

Distribution of NUTS3 regions in terms of land area and geotype (2019)

| Pop. density | Number of NUTS3 regions | Pop. (m) | Area (km2) | Urban geotype (%) | Suburban geotype (%) | Semi-rural geotype (%) | Rural geotype (%) | Extremely rural geotype (%) |
|---|---|---|---|---|---|---|---|---|
| Above 500 inh/km2 | 295 | 168.36 | 137,810 | 39.9% | 36.4% | 11.4% | 7.9% | 4.3% |
| 100 to 500 inh/km2 | 559 | 224.48 | 1,212,845 | 5.8% | 33.3% | 35.9% | 21.5% | 3.5% |
| 50 to 100 inh/km2 | 293 | 92.30 | 1,270,298 | 2.8% | 16.4% | 23.1% | 48.8% | 9.0% |
| 10 to 50 inh/km2 | 184 | 36.33 | 1,257,124 | 2.6% | 15.3% | 3.9% | 43.9% | 34.4% |
| Up to 10 inh/km2 | 17 | 2.56 | 587,219 | 1.0% | 15.1% | 0.0% | 4.7% | 79.1% |
| Total | 1,348 | 524.03 | 4,465,295 | 4.4% | 21.1% | 17.8% | 32.9% | 23.7% |

*Source: Own calculations from Eurostat data*

Table 2 shows the disparity between population density and size, in which the land area of rural geotypes (semi-rural, rural and extremely rural) in the EU-28 accounts for a 75% share of the total, while urban and suburban geotypes account for the remaining 25% of the total land size, but concentrate 75% of the population. This table offers another example of the value of breaking up NUTS3 areas into different geotypes. For instance, for NUTS3 areas that are generally regarded as simply a suburban geotype, with a relatively high population density of between 100 and 500 inhabitants per square kilometre, more than 60% of the land size in reality belongs to a rural geotype.

### 4.1.2. High-speed broadband coverage

Any new deployment of high-speed broadband networks departs from the existing fixed and wireless broadband connectivity. According to the framework above, this information is required at NUTS3 level and per geotype. To gather this information, a number of assumptions are essential, as explained below.

Firstly, data on coverage at 100 Mbps is assumed to be equivalent to the (potentially overlapping) existing deployment of FTTH, FTTB, FTTC + advanced DSL[20], and DOCSIS 3.x[21]. Secondly, in the extremely rural geotype, LTE availability is considered to be an option offering non-guaranteed 100 Mbps coverage, as discussed in the literature review (Chiha et al., 2020; Ioannou, Katsianis, & Varoutas, 2020; Ovando, Pérez, & Moral, 2015). Thirdly, regarding targets associated with an access speed of 1 Gbps, the technologies considered as a departure point are FTTH, FTTB, and DOCSIS 3.1 or superior, all of which require upgrades to comply with the relevant EGS targets. Finally, regarding 5G, we consider that the departing point is LTE; that is, that all LTE sites will be upgraded (or are being upgraded as of 2020) to 5G in addition to the deployment of new cells.

---

[20] By 'advanced DSL' we refer to technologies that fully or partially reuse existing copper infrastructure, such as G.fast over copper and bonding and vectoring techniques, as these can reach hundreds of Mbps at distances of up to 1 Km. Currently, only the UK and Switzerland are betting on this type of technology. According to Point Topic: "BT and Swisscom are leading G.fast deployment in Europe. In the UK, BT has committed to covering 10 million premises by 2020. The operator uses FTTC (fibre to the cabinet) and offers download speeds of up to 330Mbps. Swisscom has deployed FTTS (fibre to the street)- based G.fast services to 70,000 premises, bringing fibre closer to the customer. As a result, the Swiss incumbent offers maximum speeds of up to 500Mbps. Meanwhile German alternative operator M-net has rolled out G.fast in two districts of Munich. Like BT, it offers maximum theoretical download speeds of up to 300Mbps." See http://point-topic.com/g-fast-broadband-premises-3m/

Based on these assumptions, the initial data for coverage at country level can be found in the European Commission (Digital Agenda – Digital Economy and Society Index databases) for 2017 and 2019. Regional-level information on coverage can be obtained from IHS Markit, Point Topic, & Omdia (2020), which uses data for 2019 to place each NUTS3 region into one of five intervals: 0%-35%, 35%-65%, 65%-95%, 95%-100% and 100% coverage. Since these intervals are overly broad, and in order to arrive at a more precise figure, we assume that within each of the above intervals a given NUTS3 region has a level of coverage of each technology that is directly proportional to the population density. This estimation is derived from the hypothesis of "logical market behaviour", where networks are deployed first in the most populated areas. In addition, we ensure that the aggregated coverage of all NUTS3 regions in a particular country matches the data at national level, to guarantee the overall consistency of the model.

Using these assumptions, the aggregated results for all the NUTS3 regions in the EU-28 are displayed in Table 3. The results for networks that are able to provide 30 and 100 Mbps follow the expected pattern, with higher coverage in the more densely populated regions. However, for 1 Gbps, although the coverage follows the above pattern for every country, the disparity between these coverage levels (see Fig. 1) means that this effect is disguised when these figures are aggregated across Europe.

---

[21] According to the EC report on broadband (IHS Markit & Point Topic, 2018), the category of "coverage by broadband network/s capable of realistically achieving actual download speeds of at least 100 Mbps encompassed FTTP and DOCSIS 3.0 cable broadband access technologies. In cases where vectoring is applied to VDSL2 technology and speeds reach 100 Mbps and higher download speeds, VDSL with vectoring was asked to be included in this category. However, as not all connections utilizing these technologies can achieve 100 Mbps actual download speeds (for example, in the case of fibre to the building (FTTB) connections included in the FTTP category in-building wiring can pose significant constraints on achievable end-user broadband speeds), respondents were asked to exclude those connections from their answers."

**Table 3**

Coverage of high-speed broadband networks by population density

| Pop. density | Number of NUTS3 regions | 30 Mbps | 100 Mbps | 1 Gbps |
|---|---|---|---|---|
| Above 500 inh/km2 | 295 | 88.9% | 73.8% | 32.2% |
| 100 to 500 inh/km2 | 559 | 82.4% | 68.5% | 33.8% |
| 50 to 100 inh/km2 | 293 | 69.6% | 55.9% | 27.1% |
| 10 to 50 inh/km2 | 184 | 66.6% | 50.9% | 31.8% |
| Up to 10 inh/km2 | 17 | 64.8% | 42.5% | 30.8% |

*Source: Own estimation*

## 4.2. Technical path to meet EGS objectives

Another assumption for any estimation model related to EGS targets is that these targets need to be translated into practical objectives for broadband coverage. The next step is therefore to discuss how and which technologies will (or should) be deployed to achieve the desired level of coverage. Since there are different possible combinations of technologies depending on the targets and geotypes, a baseline and several alternative scenarios need to be discussed and outlined.

### 4.2.1. Targets for high-speed broadband coverage

The Intermediate 5G Connectivity objective (referred to as Target 1 or T1 in this document) calls for 5G to be available in at least one major city in each Member State. A reasonable assumption would be to select the NUTS3 region corresponding to the capital city of each Member State.

Next, the 5G Connectivity objective (referred to here as Target 2 or T2) is that all urban areas and all major terrestrial routes should have uninterrupted 5G coverage. To achieve this, data on motorways and national roads first need to be obtained from Eurostat. However, there are several different definitions of a national road. In order to err on the safe side, we select the maximum lengths among E-type roads, motorways and TEN-T networks for each country. Railway data can be obtained from Eurostat, and also from the World Bank (see Table 4). Data on both roads and railways still need to be geographically distributed among geotypes at the regional level. We assume a distribution in proportion to the land size of each geotype within each NUTS3 region.

**Table 4**

National roads and railways (km) in the EU

| Country | Roads (km) | Railways (km) | Country | Roads (km) | Railways (km) |
|---|---|---|---|---|---|
| Austria | 2,248 | 5,491 | Italy | 8,809 | 16,788 |
| Belgium | 2,026 | 3,602 | Latvia | 202 | 1,860 |
| Bulgaria | 2,953 | 4,029 | Lithuania | 1,652 | 1,911 |
| Croatia | 2,251 | 2,604 | Luxembourg | 161 | 275 |
| Cyprus | 272 | 0 | Malta | 109 | 0 |
| Czech Republic | 2,636 | 9,564 | Netherlands | 2,756 | 3,058 |
| Denmark | 1,554 | 2,131 | Poland | 5,500 | 19,132 |
| Estonia | 1,350 | 1,161 | Portugal | 3,065 | 2,546 |
| Finland | 5,229 | 5,926 | Romania | 6,200 | 10,774 |
| France | 12,797 | 28,364 | Slovakia | 1,524 | 3,206 |
| Germany | 12,996 | 38,466 | Slovenia | 773 | 1,209 |
| Greece | 4,831 | 2,240 | Spain | 15,444 | 16,167 |
| Hungary | 2,348 | 7,811 | Sweden | 6,745 | 10,882 |
| Ireland | 2,258 | 1,931 | United Kingdom | 6,926 | 16,253 |

*Source: Own estimations from Eurostat and World Bank data*

The Gigabit Connectivity objective (referred to as Target 3 or T3 in this document) aims to provide 1 Gbps connectivity for all main socio-economic institutions and digitally intensive enterprises. Assumptions for this target relate to the numbers of companies and institutions. Since there are no obvious answers to this question, a range of options has been developed to cover all possible situations. Thus, in T3.1, "all enterprises" are included, meaning 28.5 million companies and institutions. Although there is a list of suggested institutions in the EGS document, this option assumes that in 2025, all companies will have become digitally intensive in order to stay competitive, and hence will require this type of connectivity. T3.1 therefore includes the entire base of 27.5 million companies and one million institutions (local authority buildings, hospitals, schools, libraries, museums, sites of cultural interest, post offices and police stations) in the EU. However, this is four million lower than the figure used in the largest scenario for T3 by Analysys Mason (2016). For T3.2, only "5m enterprises and institutions" are considered, which is simply an intermediate option for convenience, while T3.3 considers "1m enterprises and institutions". This is similar to the smallest scenario in Analysys Mason (2016)[22]. Data on the numbers of enterprises and institutions can be taken from Eurostat.

Finally, the Rural Connectivity objective (referred to as Target 4 or T4 in this document) states that all European households, rural or urban, should have access to Internet connectivity offering a downlink of at least 100 Mbps, upgradable to Gigabit speed. For this target, a reasonable assumption is a minimum speed at 1 Gbps.

---

[22] This scenario includes 486,000 SMEs (50-249 employees) + 210,000 local authority buildings + 110,000 hospitals + 210,000 primary and secondary schools.

### 4.2.2. Choice of technologies

Assumptions related to the choice of high-speed broadband technologies to meet EGS targets are only strictly required for Gigabit Connectivity (T3) and Rural Connectivity (T4), since in the case of T1 and T2, 5G technology is already specified as part of the targets, and the only choice relates to the level of quality, as discussed below.

For T3 and T4, a "market assumption" is a continuation of the strategies already followed by incumbent and alternative providers in each country. As shown in Table 5, some countries in the EU have opted for FTTH, others for FTTB/C and still others for a combination of both technologies, depending on the geotype. This is an example of path dependency in broadband network deployment (Lemstra & Melody, 2015) and it would be extremely costly to change paths.

**Table 5**
Technology choices for the deployment of FTTx in the EU

| Dominant technology | Countries |
| --- | --- |
| FTTH | Croatia, Cyprus, Estonia, France, Hungary, Ireland, Latvia, Lithuania, Luxembourg, Malta, Portugal, Romania, Slovakia, Slovenia, Spain, Sweden |
| FTTB/C | Austria, Belgium, Bulgaria, Czech Republic, Denmark, Germany, Greece, Italy |
| FTTH in urban geotypes and FTTB/C in the rest | Finland, Netherlands, Poland, United Kingdom |

*Source: Own estimations from publicly available information at country level*

In addition, the comparative situation of cable vs. fibre deployment requires particular attention in order to avoid overlaps in deployment. In fact, according to (Queder (2020a), public policy makers should assess broadband markets separately for cable and non-cable regions, to reflect the dynamics of competition between cable operators and operators investing in FTTx networks. Table 6 summarizes the dominance of each type of technology across the EU. This information can be used to select the portion of the existing network deployment (FTTx, DOCSIS) that needs to be upgraded in order to reach the corresponding objectives at a lower cost than that of a new full deployment.

**Table 6**

Fibre vs. cable deployments in the EU (2020)

|  |  | **Fibre- FTTP** | | | |
|---|---|---|---|---|---|
|  |  | <10% | 10-25% | 25-50% | >50% |
| **Cable - DOCSIS** | <10% | Greece** | Italy*** |  |  |
|  | 10-25% |  |  |  |  |
|  | 25-50% | Ireland, United Kingdom** | Croatia, Poland | Czech Republic**, Finland, France | Latvia, Lithuania, Romania, Slovakia, Spain, Sweden |
|  | >50% | Belgium, Germany* | Austria*, Malta | Bulgaria, Hungary, Netherlands | Cyprus, Denmark, Estonia, Luxemburg, Portugal, Slovenia |

*FTTC between 10 and 25%, ** FTTC between 25 and 50%, *** FTTC more than 50%*

*Source: Own estimations from publicly available information at country level*

### 4.2.3. Technology scenarios

Table 7 summarizes the assumptions regarding technology choices for each target across geotypes. These are discussed below.

**Table 7**

Scenarios for the deployment of high-speed broadband technologies[23]

| Geotype | T1 (Intermediate) / T2 (5G) | | T3 (Gigabit) | T4 (Rural 1 Gbps) | |
|---|---|---|---|---|---|
|  | T1.1/T2.1 | T1.2/T2.2 | T3.1/T3.2/T3.3 | T4.1 | T4,2 |
|  | Guaranteed quality | Nominal quality | No. of digital intensive companies | Extremely rural wireless | Rural wireless |
| Urban | 5G | 5G | FTTH | FTTH | FTTH |
| Suburban | 5G (just for roads and railways) | | FTTH | FTTH | FTTH |
| Semi-rural |  | | FTTH | FTTH | 5G |
| Rural |  | | FTTH | FTTH | 5G |
| Extremely rural |  | | FTTH | 5G | 5G |

*Source: Own model*

For Intermediate Connectivity (T1) and 5G Connectivity (T2), there are two possible choices depending on the 5G quality: T1.1 / T2.1 "guaranteed quality", and T1.2 / T2.2 "nominal quality". This is expected to cover the range of relevant situations.

In the case of Gigabit Connectivity (T3), the choice of technology is between FTTH prepared for 1 Gbps and DOCSIS3.1[24] (depending on the situation in each country) across all geotypes. There are different options for T3 depending on the number of digitally intensive companies, as explained in the previous sub-section.

---

[23] In the case of T3 and T4, FTTH prepared for 1 Gbps and DOCSIS3.1.

[24] Several sources highlight the possibility of DOCSIS3.1 being able to provide 1 Gbps (Taga & Peres, 2018). In fact, some countries such as Germany have started the commercialization of this technology; see for instance https://www.vodafone.com/news/press-release/vodafone-germany-gigabit-investment-plan

For Rural Connectivity (T4), there are two different types of choice. In the first case, for T4.1 "extremely rural wireless", 5G is only considered for the extremely rural geotype, since it would be very costly and hence improbable that fibre-based networks would be deployed before 2025. In the second, T4.2 "rural wireless", 5G is extended to the three rural geotypes due to the cost advantages of deployment. Another assumption for T4, in line with existing market practices, is that the currently installed base for FTTH is not yet capable of 1Gbps and should therefore be updated.

These choices are merged into three different scenarios, as follows:

- Baseline investment (Baseline scenario): T1.2 "nominal quality", T2.2 "nominal quality", T3.1 "all enterprises", T4.1 "extremely rural wireless", including DOCSIS update.
- Maximum investment (Max scenario): T1.1 "guaranteed quality", T2.1 "guaranteed quality", T3.1 "all enterprises", T4.1 "extremely rural wireless", without DOCSIS update.
- Minimum investment (Min scenario): T1.2 "nominal quality", T2.2 "nominal quality", T3.3 "1m enterprises", T4.2 "rural wireless", including DOCSIS update.

## 4.3. Estimating the EGS investment gap

An estimation of the EGS investment gap first requires us to establish assumptions for the deployment costs for each technology, geotype and country, and then to quantify the gap, i.e. the investment needed to reach the targets based on the existing network coverage.

### 4.3.1. Deployment costs

The deployment costs for each technology and geotype can be obtained from the relevant academic literature (Feijóo et al., 2018; Feijóo & Gómez-Barroso, 2013; Han, Sung, & Zander, 2013; Oughton & Frias, 2016; Ovando, Pérez, & Moral, 2015; Tselekounis & Maniadakis, 2012), analysts' reports (Analysis Mason & Tech4I2, 2013; IDATE, 2013; Point Topic, 2013), studies from institutions (EIB, 2011; FCC, 2010; FTTH Council Europe, 2017; Enck & Reynolds, 2009), and industry references (5G-PPP Automotive Working Group, 2018; Vodafone, 2017)[25].

Since these costs are not necessarily congruent, they need to be merged into a single figure for each technology and geotype. The deployment costs first need to be transformed to 2019 costs using the corresponding index[26]. Next, for each type of technology and geotype, a combined deployment cost is calculated as a weighted average of all the available figures from the references, giving pre-eminence to data from specific projects at the local or municipal level, and then, in decreasing order of relevance, to costs for NUTS3 regional deployments, NUTS2 regional deployments, country deployments, and finally EU-wide deployments. This approach aims to capture the reliability of costing data by giving a higher weight to more reliable data (calculated from specific deployments), and a lower weight to less precise data (coming from EU general deployments). This heuristic procedure has also the advantage of incorporating new data as soon as they are available. Next, to include the effect of labour, it was considered that 70% of the deployment cost on average originates from civil works. Thus, this part of the cost can be adapted to comparative labour costs in each country using 2019 data from Eurostat[27]. In the final step, a factor representing the preparedness of new broadband deployments at country level was included. This factor considers geographic elements (such as whether the country is mountainous), sociodemographic elements (the predominant types of buildings), and local and national regulation elements (sharing of infrastructure availability, municipal licences and permissions). Each element is weighted based on its perceived influence,[28] and the result per country is a simple sum of the three components considered. As a result, the preparedness factor can affect deployment costs in the range ±30%. In a similar vein, following the FTTH Council, the savings opportunities are linked to the measures proposed in the Directive on Broadband Cost Reduction, such as the reuse of passive infrastructures and infrastructure sharing of up to 12% (FTTH Council Europe, 2017). See Appendix B at the end of the paper for a summary of the preparedness index.

---

[25] Figures for the update from current FTTH to 1 Gbps FTTH are drawn from discussions with main European operators.

[26] In addition to inflation, this index also includes the decrease in equipment prices due to potential increases in technology efficiency, derived from the authors' evaluation of vendor data.

It is important to note that typically estimated costs are based on the capital expenditure[29] required for network rollout to pass subscribers' premises, and do not include elements related to the core, backbone, transport, aggregation networks or customer premises equipment. In this paper, costs are expressed in euros at 2019 prices.

Table 8 below summarises the average EU deployment costs obtained from the steps described above. In the case of 5G Connectivity (T5) for roads and railways, we have used the capex per km, averaged over all geotypes. Road capex usually includes the network equipment needed for automated driving based on 5G. It is assumed that fibre for railways still needs to be deployed over 75% of their length, while fibre deployment along major roads is assumed to be still required for only 50% of their length, assumptions that are also made in other models (Analysis Mason, 2016).

---

[27] See labour cost levels by NACE Rev. 2 activity at http://appsso.eurostat.ec.europa.eu/nui/show.do?dataset=lc_lci_lev&lang=en

[28] Negative impact: -10%; neutral impact: 0%; positive impact: +10%.

[29] Hence, operating expenses (opex), maintenance of assets and WACC are not included in the model.

**Table 8**
2019 EU average deployment costs (€ per premise passed)[30]

|  | Urban | Suburban | Semi-rural | Rural | Extremely rural |
|---|---|---|---|---|---|
| FTTH | 561 | 1,376 | 2,032 | 2,633 | 6,783 |
| FTTB | 416 | 838 | 1,375 | 2,134 | 2,467 |
| FTTC | 283 | 476 | 816 | 1,380 | 1,549 |
| Upgrade to FTTH from FTTB | 188 | 643 | 813 | 870 | 4,836 |
| Upgrade to FTTH from FTTC | 321 | 1,005 | 1,372 | 1,455 | 5,754 |
| Upgrade to FTTH prepared for 1 Gbps from 100 Mbps FTTH | 112 | 275 | 406 | 527 | 1,357 |
| Upgrade to DOCSIS 3.1 from DOCSIS 3.0 | 80 | 196 | 290 | 375 | 967 |
| 5G guaranteed quality | 712 | 930 | 1,131 | 1,692 | 7,088 |
| 5G nominal quality | 444 | 565 | 687 | 871 | 1,330 |
| 5G for railways nominal quality (€/km) |  |  | 35,000 |  |  |
| 5G for railways guaranteed quality (€/km) |  |  | 55,000 |  |  |
| 5G for roads nominal quality (€/km) |  |  | 95,000 |  |  |
| 5G for roads guaranteed quality (€/km) |  |  | 115,000 |  |  |

*Source: Own calculations from a meta-analysis of the relevant literature, see above*

### 4.3.2. Investment gap quantification

Beyond the obvious calculations, an estimation of the investment gap in relation to meeting the EGS objectives requires some subtle assumptions (which are not made explicit in some of the available models) about the coverage gap for each NUTS3 region, the difference between the initial (existing) coverage and the target coverage, its conversion into the number of premises and missing connections, and, finally, the multiplication by the corresponding cost for the relevant technologies and the geotypes.

Four assumptions are required. The first relates to how new deployments take place, since the order in which geotypes and regions are covered affects the result in 2025, another example of path dependency. A simple assumption would be to again follow the market logic, which implies that the more densely populated areas are covered first. The second relates to whether cable or copper play any role in the new fixed network deployments. Here, based on previous discussions and market developments, it is supposed that all new fixed network deployments are based on fibre technologies. The third assumption relates to the issue of how to disentangle overlaps between technologies. For this, the best possible point of departure is assumed, i.e., the highest broadband footprint, which therefore describes the best possible scenario in terms of the investment gap. Finally, a logical assumption is that technologies that have already been deployed are re-used whenever possible and are economically more convenient; see for instance (Cambini & Jiang, 2009) for a similar approach.

---

[30] FTTH-GPON has been used for cost calculations in preference to FTTH P2P, due to its prevalence in the EU. In addition, the cost for the 5G spectrum is not considered. The costs for 5G in roads and railways considered here (in €/km) are common to all geotypes.

A further adjustment is needed in the case of enterprises and institutions, based on the hypothesis that covering these is more expensive than solely covering households, due to the fact that enhanced connectivity is required. A simple approach is to classify enterprises according to their size, using statistics from Eurostat, and to assume that enterprises with 0 to 9 / 10 to 19 / 20 to 49 / 50 to 249 / >250 employees require the equivalent of 2 / 5 / 11 / 50 / 100 household connections, i.e., roughly one Gbps connection for every five employees.

### 4.3.3. Planned investments from market players

The final assumption made in this methodology is the determination of which parts of the high-speed broadband investment gap will be covered by which operators, following regular market dynamics. Note that the paper does not consider any potential hindrance to broadband deployments derived from the demand side behaviour. Therefore, it is assumed that the level of broadband adoption is fully compatible with standard investment estimations from both market players and public support plans. The authors acknowledge that this is a simplistic assumption, however helpful to estimate the investment adoption gap.

Following a linear extrapolation based on an average of the investment records for previous years, it is expected that operators across the EU will provide funding of 10.4 b€ per year for fixed high-speed broadband networks (Feijóo et al., 2018). This figure falls between other previous estimations of 9.6 b€ (Analysis Mason & Tech4I2, 2013) and 14.1 b€ (Bock & Wilms, 2016). Note that although the operators are expected to invest 62.4 b€ in high-speed fixed networks up to 2025, according to the results presented in this paper, about one-third of this investment is not expected to contribute to decreasing the fixed high-speed broadband gap in the EGS calculation, as it will take place in areas where network investments from the same or similar type of operators in alternative technologies already exist, such as FTTH deployments in areas with DOCSIS or FTTC-VDSL coverage. Hence, only 41.6 b€ are accounted for in the calculation of the EGS investment gap. A previous study from Analysys Mason (2016) established a very similar figure for high-speed fixed network investment from operators, with an estimate of 81.2 b€ for the period 2018-2025.

In the case of wireless technologies, according to GSMA: "There is little guidance on 5G operator mobile capex [...] ultimately, it will depend on a number of factors including the model selected for network deployments, the targeted network coverage, the range of spectrum bands in use, and the availability of fibre infrastructure and nationwide LTE networks". Due to this uncertainty, a standard assumption involves a gradual rollout path. In fact, indications from Chinese mobile operators who have already started deployment are that 5G investment will follow a more gradual route, over a longer period than 4G (roughly seven years, from 2018 to 2025). Along similar lines, Japanese operators claim that the deployment of 5G will not lead to a significant spike in capex. For these reasons, we use the annual average figures for EU operators' capex on wireless networks for the period 2018 to 2020, which gives a figure of 22 b€ per year (GSMA, 2017).

Note also that the figure of 32.4 b€ used here for the total investment from operators per year is lower than the EC estimates of 36 b€ per year (European Commission, 2016).

## 5. Results and discussion: Investment required to meet EGS targets

This section introduces the estimation results from applying our methodology and the assumptions described above for each NUTS3 region across the EU, and then aggregates these results at country level. All targets are presented both individually and also in combination to fulfil the EGS objectives as a whole. For the latter, and as explained previously, we consider three aggregated scenarios (baseline, maximum, minimum). Some remarks on the NUTS3 level results across EU, a deeper analysis of the evolution of the investment gap from 2017 to 2019, and various comparisons with the results of other studies complete the section.

Tables 9 and 10 display a summary of the results for the EU-28 and EU-27 (without the UK), respectively. Note that some data are presented for premises (defined as households plus enterprise locations) to provide a more realistic view of the broadband investment gap.

**Table 9**
Summary of EGS investment needs (b€) in the EU-28

| EU-28 | Baseline | Max | Min |
| --- | --- | --- | --- |
| T1 (Intermediate connectivity – major city) | 14.7 | 23.9 | 14.7 |
| T2A (5G connectivity – urban areas) | 45.1 | 72.3 | 45.1 |
| T2B (5G connectivity – transport paths) | 16.9 | 23.0 | 16.9 |
| T2 once T1 is achieved | 50.3 | 76.5 | 50.3 |
| T3 (Gigabit Connectivity) | 91.8 | 94.2 | 4.1 |
| T4 (Rural Connectivity) | 156.4 | 159.9 | 85.1 |
| **EGS (premises)** | 221.4 | 260.3 | 150.2 |
| **EGS (premises + companies)** | 294.9 | 335.8 | 153.9 |

*Source: Own calculations*

**Table 10**
Summary of EGS investment needs (b€) in the EU-27

| EU-27 | Baseline | Max | Min |
|---|---|---|---|
| T1 (Intermediate connectivity – major city) | 12.8 | 20.9 | 12.8 |
| T2A (5G connectivity – urban areas) | 36.8 | 59.0 | 36.8 |
| T2B (5G connectivity – transport paths) | 15.7 | 21.3 | 15.7 |
| T2 once T1 is achieved | 42.7 | 64.6 | 42.7 |
| T3 (Gigabit Connectivity) | 83.8 | 85.5 | 3.8 |
| T4 (Rural Connectivity) | 140.2 | 142.9 | 71.0 |
| **EGS (premises)** | **195.7** | **228.4** | **126.5** |
| **EGS (premises + companies)** | **262.5** | **296.7** | **129.9** |

*Source: Own calculations*

## 5.1. Intermediate Connectivity – 5G available in a major city in each Member State by 2020 (T1)

The investment needed to meet the Intermediate Connectivity objective (T1) for the EU-28 amounts to 14.7 b€ in the baseline scenario. Choosing a higher quality for 5G provision increases this investment to 23.9 b€.

At the country level, the investment simply depends on the size and population of the capital region of each Member State, with France (3.6 b€), the UK (1.9 b€) and Spain (1.3 b€) leading the baseline for T1 (see Fig. 8).

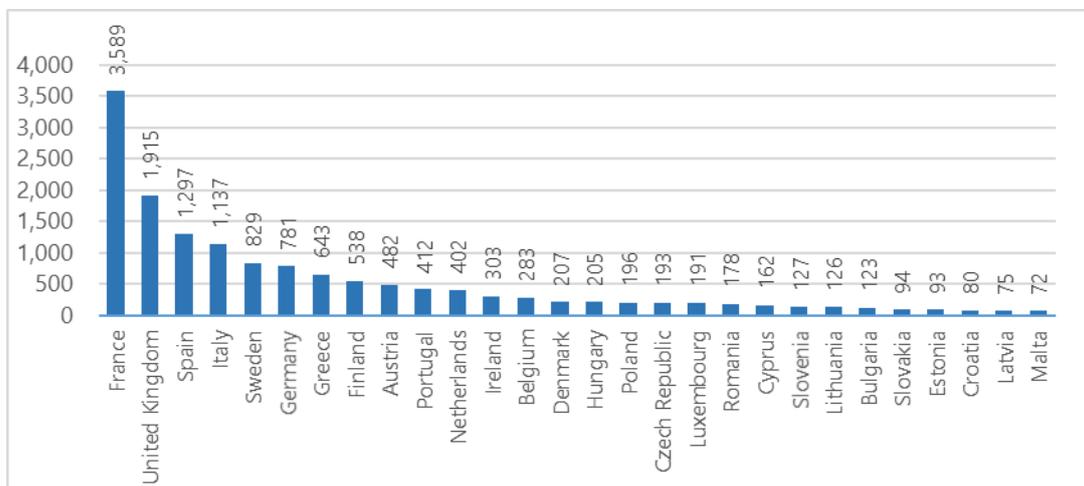

**Fig. 8.** Investment needed to complete the availability of 5G in the capital city of each EU Member State by 2020 (Intermediate Connectivity - T1 baseline: T1.2 "nominal quality") in m€

## 5.2. 5G Connectivity – uninterrupted 5G coverage in all EU urban areas and along major terrestrial transport paths by 2025 (T2)

The investment needed to fulfil the 5G Connectivity objective (T2) amounts to 62.0 b€ for the EU-28 in the baseline scenario. Choosing a higher quality for 5G provision increases the investment to 95.3 b€.

This objective can be divided into two parts: (i) 5G coverage in urban areas, and (ii) 5G coverage along major terrestrial paths. The baseline scenario for urban areas requires an investment of 45.1 b€, while terrestrial paths including railways will need 16.9 b€. In the case of higher-quality provision of 5G, the urban expenses would increase to 72.3 b€, and for terrestrial paths this would be 23.0 b€ (see Fig. 9). In a study of ubiquitous connectivity, Analysys Mason (2016) provides a figure of 5.2 b€ for railways, which rises to 6.7 b€ when motorways are included and 28 b€ when state roads are added — the option favoured by the EC (European Commission, 2016) — and finally 103 b€ if provincial roads are also considered.

If the 5G Connectivity (T2) target is met after the deployment of the Intermediate Connectivity objective (T1), then the total investment decreases to 50.3 b€ in the baseline scenario and to 76.5 b€ for the higher quality scenario.

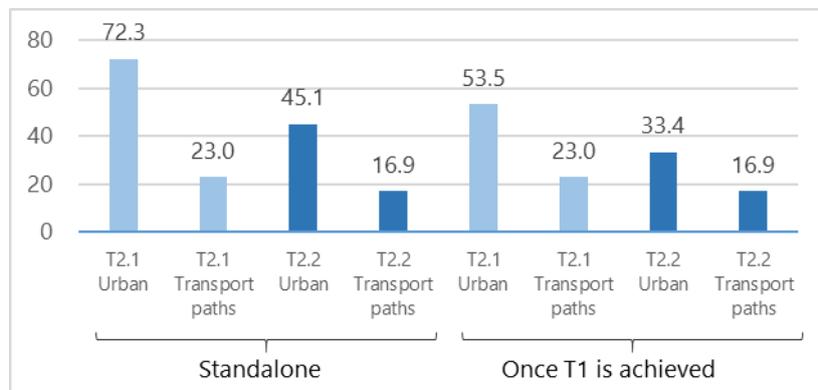

**Fig. 9.** Investment needed to provide uninterrupted 5G coverage in all EU urban areas and along major terrestrial transport paths by 2025 (5G Connectivity - T2) in b€

At the country level, the required investment in the baseline scenario is highest in France (10.8 b€), the UK (9.5 b€), and Germany (8.8 b€). Together, these account for almost 50% of the total investment gap. Lower investment is required in countries with smaller territories and populations, such as Malta, Luxembourg, Cyprus, Latvia, Estonia, Slovenia and Lithuania (see Fig. 10).

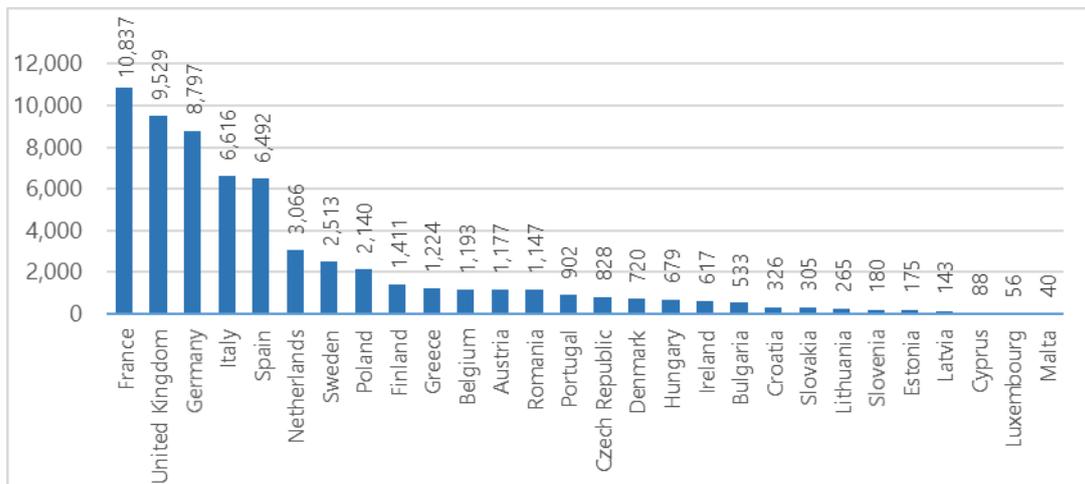

**Fig. 10.** Investment needed to provide uninterrupted 5G coverage in all EU urban areas and along major terrestrial paths in each Member State by 2025 (5G Connectivity - T2 baseline: T.2 "nominal quality") in m€

## 5.3. Gigabit Connectivity for all EU socio-economic drivers by 2025 (T3)

The investment needed to complete the Gigabit Connectivity objective (T3) for the EU-28 amounts to 91.8 b€ in the baseline scenario. If only five million EU companies and institutions are considered, this investment is reduced to just 18.4 b€, while limiting the number to one million enterprises and institutions causes the investment to be as little as 4.1 b€ (see Fig. 11).

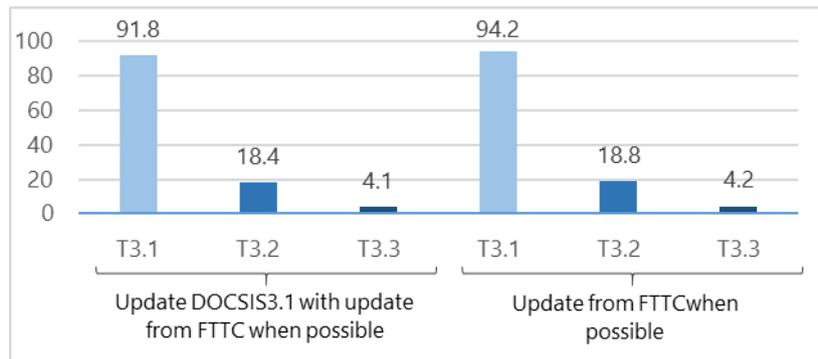

**Fig. 11.** Investment needed to provide 1 Gbps connectivity for all EU socio-economic drivers by 2025 (Gigabit Connectivity - T3) in b€

At the country level, the required investment is highest in France (24.4 b€), followed by Italy (16.1 b€) and the UK (8.0 b€). Together, these account for more than 50% of the total investment needed. Lower investment is required in countries with smaller sizes, such as Malta, Luxembourg, Cyprus, Latvia, Estonia, Denmark and Slovenia (see Fig. 12).

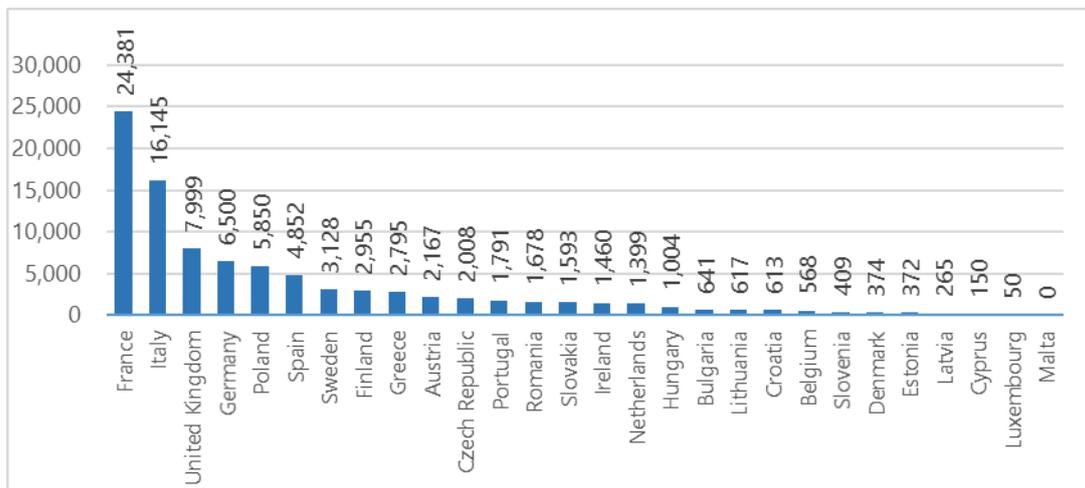

**Fig. 12.** Investment needed to provide 1 Gbps connectivity for all EU socio-economic drivers in each Member State by 2025 (Gigabit Connectivity - T3 baseline: T3.1 "all enterprises") in m€

## 5.4. Rural Connectivity – 1 Gbps connectivity for all EU households by 2025 (T4)

The investment needed to meet the Rural Connectivity objective (T4) for the EU-28 amounts to 156.4 b€ in the baseline scenario in which the upgrade of DOCSIS is considered, and to 159.9 b€ if the DOCSIS upgrade is not applied (see Fig. 13). If the wireless option is extended to include the three rural geotypes, the investment required decreases to 85.1 and 88.5 b€, with and without DOCSIS upgrade, respectively; both of these are somewhat lower than the EC's figure for a combination of fibre and wireless for this objective at 127 b€ (European Commission, 2016).

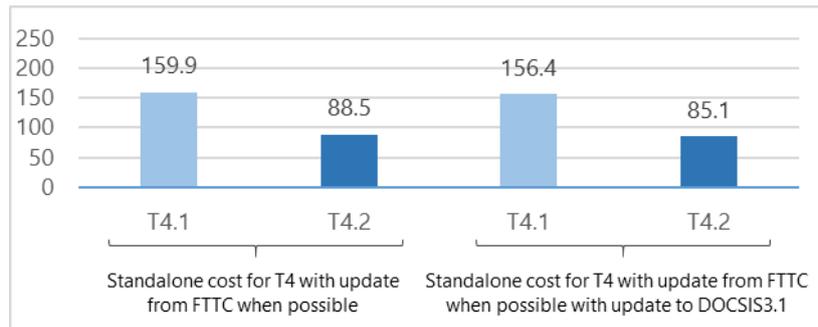

**Fig. 13.** Investment needed to provide 1 Gbps connectivity for all EU households by 2025 (Rural Connectivity - T4) in b€

At the country level, the required investment in the baseline scenario is highest in France (44.7 b€), followed by Italy (25.9) and Germany (17.1 b€). Together, these account for almost 60% of the total investment needed. Lower investment is required in countries with a smaller size and/or existing rather developed broadband networks, such as Malta, Luxembourg, Cyprus, Latvia, Estonia and Slovenia (see Fig. 14).

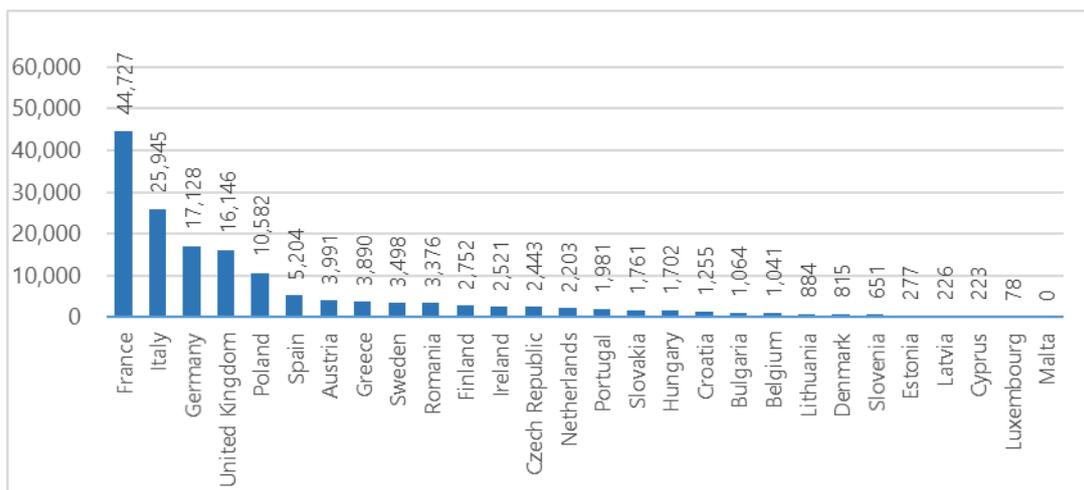

**Fig. 14.** Investment needed to provide 1 Gbps connectivity for all EU households by 2025 (Rural Connectivity - T4 baseline: T4.1 "extremely rural wireless") in m€

The T4 investment gap can be also evaluated as a function of geotype, as Table 11 shows. As expected, the gap is higher in rural areas; it is also in rural areas where the most advanced fixed technologies are missing, and in particular FTTH. In addition, while the share of the gap in rural areas is 33% of the premises vs. 67% in urban and suburban areas, the weight in terms of investment is much higher: a 68% share vs. a 32% share in urban areas. The difference between urban and rural areas is also highlighted in terms of the change in cost per premise, which is much higher in rural areas. The exception is extremely rural areas, which are comparatively low since they use 5G at nominal quality.

**Table 11**
2019 T4 baseline investment gap by geotype

| EGS (premises) | Urban | Suburban | Semi-rural | Rural | Extremely rural |
| --- | --- | --- | --- | --- | --- |
| Share of gap in number of premises | 31% | 36% | 21% | 11% | 1% |
| Share of gap in terms of investment | 6% | 26% | 38% | 29% | 2% |
| Total investment (b€) | 9.4 | 40.0 | 58.7 | 45.6 | 2.6 |
| Average investment per premise (€) | 179 | 664 | 1,629 | 2,541 | 1,296 |

## 5.5. EGS investment by 2025 (T1+T2+T3+T4)

For the overall EGS results, our analysis starts from the deployment situation in relation to high-speed broadband, as of 2019, to estimate the broadband investments needed from 2020 ("the 2020 EGS investment gaps"). Intermediate 5G Connectivity (T1) is considered first, and total 5G Connectivity (T2) is then added, taking into account the fact that T1 already covers T2 except for the non-urban geotypes contained within the capital city of each Member State. Rural Connectivity (T4) and then Gigabit Connectivity (T3) are then introduced to complete the scenario, but with two additional considerations that require certain adjustments in order to fine-tune the model: part of T3 is already covered by T4 (since the paper is based on premises connectivity rather than households), and T3 requires FTTH in the extremely rural geotype, in order to provide connectivity to digitally intensive companies.

Following the steps above, the baseline for the EGS for the EU-28 amounts to 294.9 b€, and a higher quality wireless provision for 5G would increase the investment to 335.8 b€. In a similar scenario, the EC estimated the cost of the EGS objectives for the period 2018–2025 at 443 b€. If wireless technologies are chosen for the three rural geotypes in the case of Rural Connectivity (T4), then the required EGS investment drops to 153.9 b€, while if only premises are considered (i.e. excluding the effects of Gigabit Connectivity (T3)), then the required EGS investment for premises in the baseline case is 221.4 b€.

At the country level, France (75.5 b€), Italy (45.7) and the UK (32.3 b€) account for more than 50% of the baseline investment need. Lower investment is required to fulfil the EGS objectives according to the baseline scenario in Malta, Luxembourg, Cyprus, Latvia and Estonia (see Fig. 15).

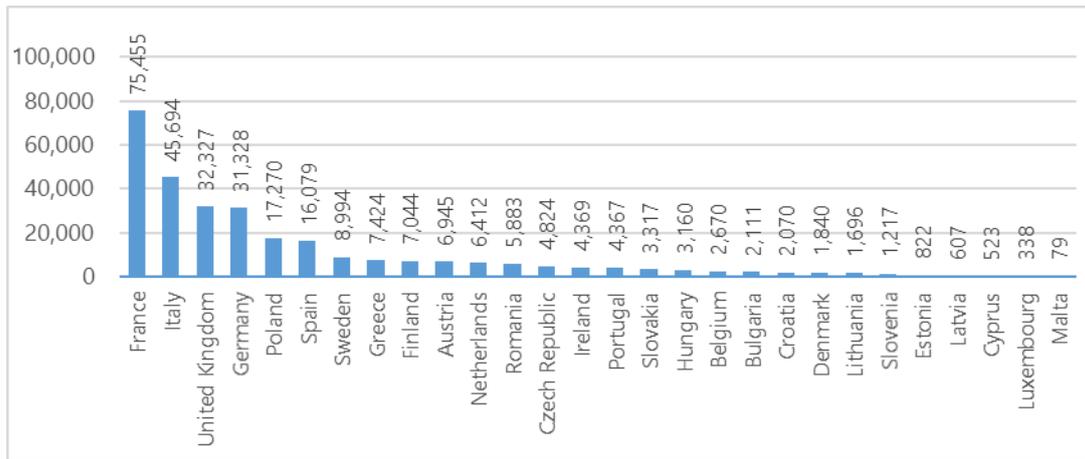

**Fig. 15.** Investment need to meet the EGS targets (T1 + T2 + T3 + T4) by 2025, by Member State (T1+T2 baseline: T1.2 + T2.2 "nominal quality", T3 baseline: T3.1 "all enterprises", T4 baseline: T4.1 "extremely rural wireless") in m€

## 5.6. NUTS3 analysis of EGS investment

The histogram in Fig. 16 shows the distribution of the NUTS3 regions according to the estimated gap in premises to complete 1 Gbps broadband coverage. This gap is lower than 50% of premises only in 163 NUTS3 regions, that account for 22% of the total EU-28 population. On the contrary, this gap is higher than 50% of premises in 1,185 regions, that concentrate 78% of the total population. The low coverage of 1 Gbps broadband networks in larger countries such as the UK and Germany is the main reason for this distribution.

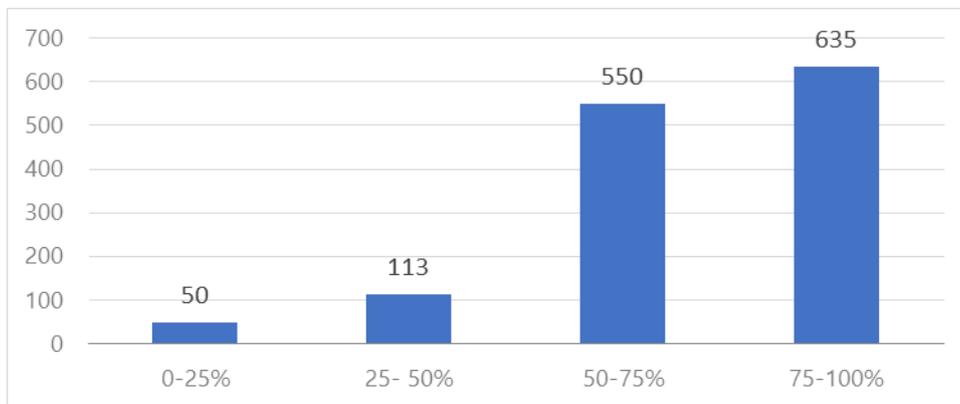

**Fig. 16.** Histogram of numbers of NUTS3 regions as a function of the EGS investment need (in % premises) in the EU-28

The investment needed to meet all EGS objectives (see Fig. 17) is fairly similar for the urban and suburban geotypes (58.8 and 64.4 b€, respectively), and somewhat higher for the semirural and rural geotypes (86.3 and 70.0 b€, respectively), where a lower number of premises compensates for the larger gap in the percentage of premises and the more expensive unitary costs for the different fixed technologies deployed. The investment required in the rural geotype, however, is significantly lower (14.3 b€) due to the reduced number of premises and the use of 5G instead of fibre for the Rural Connectivity target (T4). It should also be noted that most of the investment in the urban geotype is required to meet the Intermediate Connectivity (T1) and 5G Connectivity (T2) objectives, as they mainly affect urban areas, while the Gigabit Connectivity (T3) and Rural Connectivity (T4) objectives affect only a limited number of premises in this geotype. In the rest of geotypes, most of the investment is needed to meet T3 and T4, which are affected by the lower current penetration of high-speed broadband and consequently have larger gaps to cover.

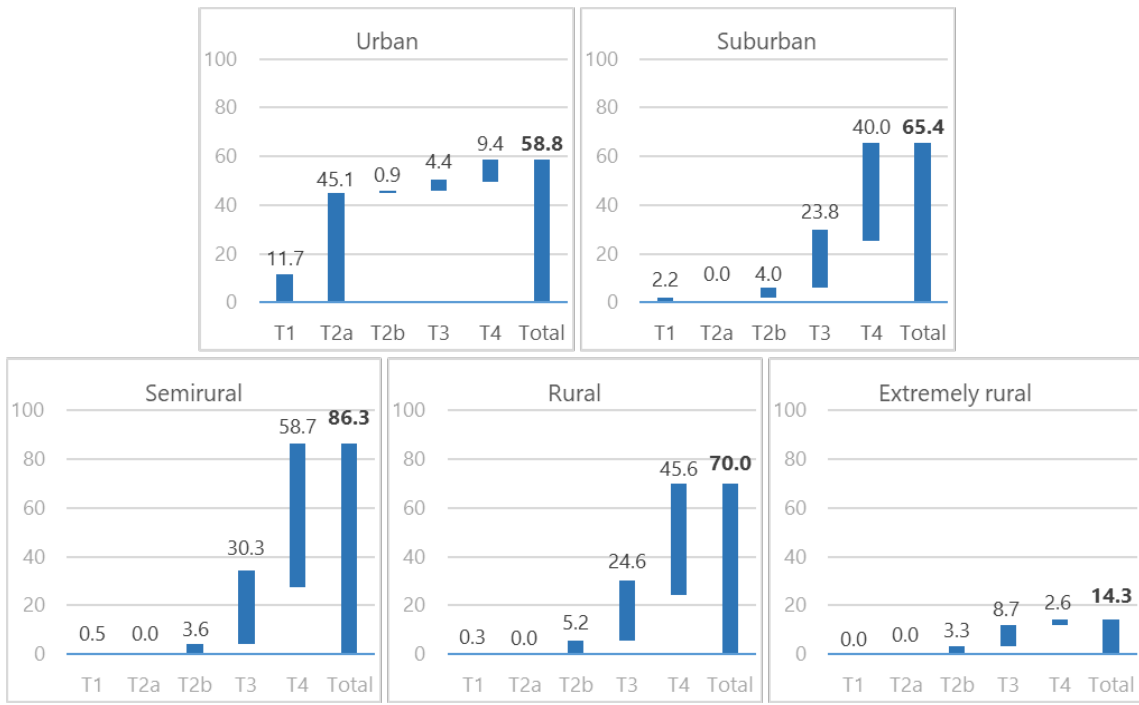

**Fig. 17.** Investment needed to meet the EGS targets (T1 + T2 + T3 + T4) by 2025 per geotype, in the baseline scenario (T1+T2 baseline: T1.2 + T2.2 "nominal quality", T3 baseline: T3.1 "all enterprises", T4 baseline: T4.1 "extremely rural wireless") in b€

## 5.7. EGS investment breakdown

It is also possible to break down the EGS investment into different components (see Table 12). Firstly, the urban–rural divide has already been investigated, with pending investment in urban and suburban areas being lower than in rural ones (42% vs. 58%), despite the 77% share of urban premises. Secondly, if companies are included in the calculations, the investment increases to 33% above the premises baseline, whereas if only households rather than premises are considered, the investment decreases by about 11%. Thirdly, with regard to cohesion regions across the EU, their share of the total investment for EGS is 33% of the total. Finally, we note that the UK accounts for about 12% of the total investment need across all scenarios.

**Table 12**

Breakdown of EGS investment needs EU28 (b€)

|  | **Baseline** | **Max** | **Min** |
|---|---|---|---|
| EGS (Total households) | 196.1 | 235.0 | 124.8 |
| EGS (Total premises) | 221.4 | 260.3 | 150.2 |
| EGS (Total premises + companies) | 294.9 | 335.8 | 153.9 |
| EGS (Households vs. premises) | 89% | 90% | 83% |
| EGS (Premises + companies vs. premises) | 133% | 129% | 102% |
| EGS (EU-27) (premises) | 88% | 88% | 84% |
| EGS (UK) | 12% | 12% | 16% |
| EGS (Urban and suburban) | 42% | 47% | 67% |
| EGS (Semi-rural, rural and extremely rural) | 58% | 53% | 33% |
| EGS (Cohesion regions) | 39% | 37% | 34% |
| EGS (Non-cohesion regions) | 61% | 63% | 66% |
| EGS (Expected operators' investment: Premises + companies) | 36% | 42% | 69% |
| EGS (Expected investment gap) | 64% | 58% | 31% |

*Source: own calculations*

The model also estimates that, once the network deployments from incumbent and/or alternative operators in the period 2020-2025 get completed, there would remain a gap of 188 b€ to achieve the EGS targets. This is a higher figure than the EC's estimate of 155 b€ (European Commission, 2016), but lower than that of Analysys Mason (2016) at 251 b€. Hence, deployment from existing operators is expected to reduce the high-speed broadband investment gap by a little more than one third of the initial total investment need, in the baseline scenario. Additional public policies, in the form of a revised regulatory framework or new subsidies, among others, would be helpful to close this gap. Note, however, that in the minimum scenario, the situation is reversed, and investment from operators makes up almost two thirds of the total investment required.

The paper assumes that the level of broadband connectivity demand from end users in the period 2020-2025 is high enough to justify both the planned network rollouts by operators and their expected return on the required investment. Authors acknowledge that demand has been impacted by COVID-19 pandemic in different directions, still not fully understood which will be the result of their combination. On one side, it causes a need for more and better broadband connectivity, due to more frequent and demanding online activities, such as work from home, online education or remote medical assistance, among many others. On the other side, a decline on the demand of connectivity might happen, due to lower spending from households, enterprises and public institutions, all affected by the contraction of the economy.

## 5.8. Evolution of the investment required and comparison with other studies

In this section, we analyse how the investment required to meet the EGS targets evolved between 2017 and 2019, and compare the results with other analyses in the literature. We follow the same methodology and make the same assumptions as discussed previously, and use two different datasets: firstly, the latest data available from IHS Markit, Point Topic, & Omdia (2020), which were used to obtain the results presented above and which refer to 2019, and secondly, the previous version of the same dataset, published two years earlier (IHS Markit & Point Topic, 2018), which refer to year 2017. The rest of the input data on demographic characteristics across countries and NUTS3 regions were obtained from Eurostat and the EC for the corresponding moments in time, to ensure consistency throughout the analysis.

As discussed in previous sections, the total investment required to meet all the EGS objectives for the EU-28 reaches 294.9 b€ in the baseline scenario (see Fig. 18), as compared with 341.8 b€ predicted two years ago and 393.2 b€ estimated by Analysys Mason (2016). The main reason for the reduction over this period is the increase in the coverage of networks that are already capable of providing 1 Gbps, from mid-2017 to mid-2019; this has reduced the gap and the investment required to meet the Gigabit Connectivity and Rural Connectivity targets (T3 and T4, respectively). A simple linear regression extrapolation gives an average yearly reduction of 24.6 b€ during the period 2016–2020. If this linear trend continues unabated (which is highly improbable due to the diminishing returns of deploying high-speed broadband networks in rural areas), there would still be an investment gap of 161.5 b€ in 2025 (the estimation made here is based on 188 b€ of pending investment for 2025, see above), and the gap would be closed in 2032.

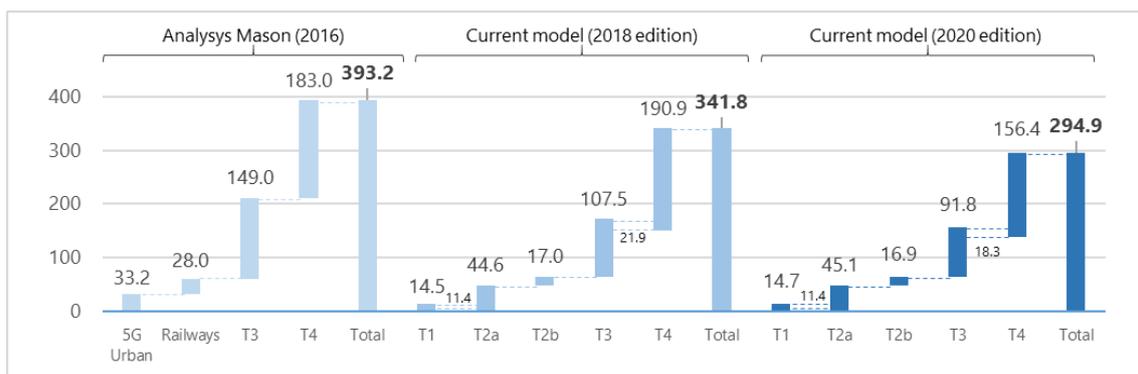

**Fig. 18.** Investment need to achieve EGS targets (T1 + T2 + T3 + T4) by 2025, in b€

The same analysis can be carried out by looking at each target individually. In the case of Intermediate Connectivity (T1), the investment reaches 14.7 b€, as compared with 14.5 b€ for mid-2017. For the 5G Connectivity target (T2), we obtain a required investment of 62.0 b€, slightly higher than the 61.6 b€ obtained for mid-2017. In both cases, the slight increases arise from the higher number of premises, which is especially noticeable in large countries like the UK and Italy, and from higher labour costs, especially in the UK.

The gap in the number of premises, however, does not change significantly from one analysis to the next, as most 5G deployments are yet to be implemented in Europe and the EC analysis does not include details of 5G coverage in the dataset.

For the next target, Gigabit Connectivity (T3), the investment amounts to 91.8 b€ in the baseline scenario, a figure that is significantly lower than the 107.5 b€ from mid-2017. The main reason for this reduction lies in the deployment of fixed networks during this period, which has increased the number of premises already covered by networks capable of providing 1 Gbps and has consequently reduced the number of premises to be covered by new deployments. The reduction is especially significant in those countries where 1 Gbps coverage is higher. The value obtained for T3 is also lower than the 149 b€ from Scenario D presented by Analysys Mason (2016), since the latter is based on 4 million more companies and an additional 10 million teleworkers and freelancers.

From looking at each country individually, we find that in the case of Malta, the investment gap is reduced to zero; as of mid-2019, the country has already achieved 100% coverage, and no further deployment is required. However, it should be noted that in some other countries, the investment gap has counterintuitively increased, which is mostly due to changes in the demographic information used to characterise the geotypes. In Lithuania, for instance, the investment has doubled, due to the difference in the distribution of premises across geotypes. At the time when the first analysis was carried out, there was only sufficient information to distribute a small part of the total base of premises across the geotypes. The new data that have become available have solved this issue and allow for the allocation of the whole base of premises in the country across the geotypes, resulting in a higher concentration in rural areas and subsequently higher costs of deployment. In some other cases, the investment has increased due to significant changes in the data provided by the EC and used as input for the analysis. For instance, for certain countries, the level of 100 Mbps coverage provided in mid-2019 is lower than the figure provided in mid-2017, leading to larger requirements in terms of network deployment and subsequently higher investment. This effect is particularly noticeable in Portugal, where the required investment has almost doubled.

For the Rural Connectivity target (T4), we estimate a gap of 156.4 b€ in the baseline scenario, a significant reduction in relation to the 190.9 b€ obtained for mid-2017. The main reasons behind this reduction, and also behind certain atypical increases for several countries, are the same as for the Gigabit Connectivity target. These results are comparable with the 183 b€ estimated by Analysys Mason (2016) in its Scenario E for Rural Connectivity, which was fully based on fibre. The FTTH Council Europe estimates full fibre deployment at between 137 and 156 b€ (FTTH Council Europe, 2017), depending on the cost savings from sharing existing infrastructures, but does not consider the extension to 1 Gbps, which would add about 30 to 50 b€ to these figures. If the wireless option is extended to include the three rural geotypes, the investment required to meet the Rural Connectivity target (T4) decreases to 85.1 b€, significantly lower than the 111.7 b€ obtained for 2018 and the 127 b€ considered by the EC (European Commission, 2016).

## 6. Conclusions

The primary conclusion of this paper is that despite their different assumptions, each model of the deployment of high-speed broadband networks in the EU agrees on the impending necessity of a very large investment to complete this coverage, and specifically to achieve the EGS targets by 2025, especially in suburban, semi-rural and rural areas.

These different estimation models also broadly coincide in terms of the amount of investment required. For the period 2018–2025, this investment gap is in the range of 370 to 395 b€ for the EU-28 at 5G nominal quality, and from 340 to 365 b€ for the EU-27. To authors' knowledge, the current paper has presented the only estimation of the investment gap for the period 2020–2025, which has been reduced to 295 b€ for the EU-28 and 260 b€ for the EU-27.

Another important conclusion of this paper is that only about one third of these figures are covered by the funding already expected from private operators, leaving a remaining investment gap in the range of 161 to 188 b€. This result demonstrates that high-speed broadband networks share some features of an economic public good, with private investors appropriating only a part of the total benefits of broadband connectivity (Gruber Hätönen, & Koutroumpis, 2014). Other studies examined in this paper also report the need for a public push to complete the deployment of high-speed broadband networks. Suggested public intervention could take the form of existing regulation, such as that affecting wholesale vs. integrated operators (Queder, 2020b), or alternative forms of regulation to foster investment in ultra-fast broadband networks, such as geographically differentiated access pricing and an obligation to co-invest or share investment costs among different operators (Abrardi & Cambini, 2019b). A study of co-investment in fibre in France suggests that there could be considerable increases in coverage as a function of the extent of infrastructure shared (Lebourges & Liang, 2018), while another in the Netherlands stresses the policy hurdles that will be encountered even if community-based deployments are fostered (Salemink & Strijker, 2018).

With regard to the specific EGS objectives, Targets 1 (Intermediate 5G Connectivity) and 2 (5G Connectivity) are less demanding in terms of investment; together, these amount to 65 b€ of the 295 b€ in the baseline scenario, as they can be met with narrower deployments of wireless technologies that are easier and less expensive to deploy. On the other hand, Targets 3 (Gigabit Connectivity) and 4 (Rural Connectivity) require a much higher investment representing 230 b€ of the 295 b€ in the baseline scenario, as they cover a much wider area and require the deployment of more expensive fixed technologies in the more costly rural geotypes. This is a finding that is common to all of the estimation models examined.

Although the investment needed to meet the EGS targets amounts to 295 b€ in the baseline scenario, this figure may also change considerably in alternative scenarios. A preference for fibre-based technologies in the extremely rural geotype combined with higher-quality 5G deployments increases the size of the investment by 14%, to 336 b€. Conversely, if wireless technologies at nominal quality are deployed in all three rural geotypes, rather than solely in the extremely rural case, the investment is reduced by half compared with the baseline scenario, to 154 b€. Thus, technology-agnostic network deployments for Target 4 (Rural Connectivity) have started to be considered in deployment models.

In practice, the performance of wireless technologies plays a key role in any analysis of future high-speed broadband deployments. We are currently at the first stages of development of 5G, and any future improvement in this technology can only improve its performance in terms of capacity and increase the opportunity to use wireless technologies instead of fibre (at least in the last stretch to the user), even in urban and suburban geotypes. This could substantially reduce the investment required to meet the more demanding objectives, such as Targets 3 and 4. It will also impact the balance between the total investment gap and the investment from market players. An aggressive substitution of fibre by 5G in the rural geotypes would reduce the need for additional funding from two thirds to one third of the total investment gap, as shown in the analysis. Wireless technologies can also be viably adapted to rural environments in the form of a fixed wireless access network, as already proven in a model for a competitive reduction of costs when using LTE (Ioannou, Katsianis, & Varoutas, 2020). If the quality provided to rural users is adjusted and wireless technologies are used in combination with satellite links, the deployment costs can be further reduced, as other estimations from deployment models attest (Chiha et al., 2020).

A more granular analysis focused on countries and NUTS3 regions shows that some will almost certainly meet the EGS targets before 2025. There is a strong path dependency for every country in this situation; the main reasons for this include leverage on past investments, the existing footprints of fibre and cable networks, current investment plans and support from the government, the (small) size of the country, the level of urbanisation and the cost of deployment, as highlighted in previous papers (Feijóo et al., 2018). From the perspective of NUTS3 areas, the results presented in this paper predict a noticeable investment gap in most of the NUTS3 regions in Europe, affecting about 80% of the total EU population, with a particular impact on rural areas.

During the period 2017–2019, the deployment of fibre-based fixed networks has contributed to reducing the gap from 342 b€ to 295 b€ in the baseline scenario presented here, i.e. about 25 b€ per year. However, the remaining gap is still challenging, and further action seems necessary to achieve or even to approach to EGS targets. The deployment of telecommunications networks is very demanding in terms of time and resources. New measures to promote public and private investments, with a particular focus on rural areas, are required to ensure the wide availability of high-speed broadband networks, and the sooner the better.

In summary, it can be seen that bottom-up estimation models, despite their apparent complexity and number of assumptions, have a proven capacity to allocate a higher geographical diversity of broadband networks and technologies, the flexibility required to reflect suitable combinations of technologies and the effects of policies and regulations, robustness, and the ability to provide sufficiently accurate figures to guide policy action at local, regional, national and supra-national levels. At the same time, the explainability power and usefulness of these estimation models at EU level would be enhanced by the availability of homogeneous and stable information on socio-demographics and the levels of coverage and take-up of broadband networks, at least at the NUTS3 level. The latter would also contribute to a more transparent market dynamics, and would improve the chances of success of any public policy action regarding high-speed broadband deployment.

**Acknowledgements**

Part of the analysis carried out in this paper is based on previous work by the authors for the European Investment Bank (EIB) on the future of high-speed broadband deployments in Europe. We would like to thank EIB for their support.

# Appendix A. Gbps network initiatives and commitments in the EU (2018)

| Country | Operator | Households (m) | Target year | Source |
|---|---|---|---|---|
| Austria | | | | |
| Belgium | | | | |
| Bulgaria | | | | |
| Croatia | | | | |
| Cyprus | CYTA | 0.2 | 2028 | https://cyprus-mail.com/2018/07/05/cyta-launches-fibre-optic-network/ |
| Cz Republic | | | | |
| Denmark | TDC | Half | 2017 | https://www.huawei.com/en/press-events/news/2016/1/Huawei-and-TDC-Group |
| Denmark | TDC | Whole | 2018 | https://www.digitaltveurope.com/2016/01/27/tdc-pledges-denmark-wide-1gbps-broadband-in-2018/ |
| Estonia | Starman | 0.05 | 2022 | https://www.zdnet.com/article/like-the-sound-of-10-gigabit-internet-everywhere-then-move-to-estonia/ |
| Finland | DNA | 0.6 | 2017 | https://www.telegeography.com/products/commsupdate/articles/2016/05/10/dna-finland-launches-gigabit-broadband/ |
| France | | | | |
| Germany | Deutsche Telekom | 0.04 | 2019 | https://www.telekom.com/en/blog/group/article/broadband-facts-versus-legends-516794 |
| Germany | Vodafone | 25 | 2022 | https://www.telcotitans.com/vodafonewatch/vodafone-expands-gigabit-connections-in-europe/2024.article |
| Greece | OTE | 1 | 2022 | https://www.telecompaper.com/news/ote-to-provide-1-mln-households-businesses-with-access-to-ftth-network--1238409 |
| Hungary | Digi | 0.7 | - | https://bbj.hu/business/digi-to-expand-1-gbps-internet-to-all-customers_148197 |
| Ireland | Siro (Vodafone&ESB) | 0.5 | 2018 | https://www.siliconrepublic.com/comms/siro-south-east-ireland-fibre-broadband-axione-obelisk |
| Italy | TIM | 2.7 | 2018 | http://www.telecomitalia.com/tit/en/about-us/business/rete-strategia.html |
| Latvia | Lattelecom | 0.492 | 2014 | https://www.computerweekly.com/news/2240214496/Latvias-Lattelecom-1Gbps-will-be-standard-by-2018 |
| Lithuania | Skynet | 0.119 | 2016 | https://www.telecompaper.com/news/skynet-claims-fastest-broadband-in-lithuania--1126430 |
| Luxembourg | LuxConnect | Whole | 2020 | https://www.luxconnect.lu/dark-fiber/ |
| Malta | Melita | Whole | 2020 | https://www.globenewswire.com/news-release/2020/02/14/1985252/0/en/Malta-s-Melita-Providing-Gigabit-Broadband-Service-Nationally.html |
| Netherlands | KPN | 3 | 2021 | https://www.overons.kpn/en/news/2020/kpn-introduceert-1-gbit-s-glasvezelinternet |
| Poland | Orange Poland | 1.7 | 2018 | https://www.telecompaper.com/news/orange-poland-upgrades-top-broadband-speed-to-1-gbps--1257856 |
| Portugal | Vodafone | 2.75 | 2016 | https://www.vodafone.com/content/index/about/policy/news/public-policy-news-releases/2015/gigabit-fibre-europe.html |
| Romania | | | | |
| Slovakia | Slovak Telekom | 0.5 | 2018 | https://www.telecompaper.com/news/slovak-telekom-offers-1-gbps-speed-over-fibre-as-add-on--1238941 |
| Slovenia | | | | https://www.telegeography.com/products/commsupdate/articles/2013/04/26/vahta-challenges-ftth-rural-reluctance/ |
| Spain | Orange | 12 | 2018 | http://blog.orange.es/adsl-fibra/lanzamos-fibra-1gbps-simetrico-12-millones-hogares-wi-fi-inteligente/ |
| Spain | Vodafone | 10.3 | 2018 | https://www.elespanol.com/economia/empresas/20180725/vodafone-podra-gbps-clientes-red-vuelta-cole/325217964_0.html |
| Sweden | Stokab | 0.4 | 2018 | http://www.ftthcouncil.eu/documents/FTTH_Council_report__FINAL_and_proofread-update-20180214.pdf |
| UK | Government project | | | https://www.gov.uk/government/news/six-areas-to-pilot-uks-fastest-broadband-as-part-of-200-million-project |
| UK | Openreach / BT | 3 | 2020 | https://www.homeandbusiness.openreach.co.uk/fibre-broadband/ultrafast-broadband/ultrafast-fibre-fttpfurl |
| UK | Virgin Media | 15 | 2021 | https://www.libertyglobal.com/virgin-media-to-bring-gigabit-internet-to-millions-of-homes/ |

## Appendix B. Network preparedness factors for the EU[31,32,33]

| Country | Geographic landscape factor | Housing factor | National / local regulation factor | Combined preparedness factor |
|---|---|---|---|---|
| Austria | 10% | 0% | -10% | 0% |
| Belgium | -10% | 0% | -10% | -20% |
| Bulgaria | 10% | 0% | 10% | 20% |
| Croatia | 0% | 0% | 10% | 10% |
| Cyprus | 0% | 0% | 10% | 10% |
| Czech Republic | 0% | -10% | 0% | -10% |
| Denmark | 0% | 0% | -10% | -10% |
| Estonia | -10% | -10% | 0% | -20% |
| Finland | 10% | 0% | -10% | 0% |
| France | 0% | 0% | 0% | 0% |
| Germany | -10% | -10% | -10% | -30% |
| Greece | 10% | -10% | 10% | 10% |
| Hungary | -10% | 0% | 10% | 0% |
| Ireland | -10% | 10% | -10% | -10% |
| Italy | 10% | -10% | 10% | 10% |
| Latvia | 0% | -10% | 0% | -10% |
| Lithuania | 0% | -10% | 0% | -10% |
| Luxembourg | -10% | 0% | -10% | -20% |
| Malta | 0% | -10% | 0% | -10% |
| Netherlands | 0% | 10% | -10% | 0% |
| Poland | -10% | 0% | 10% | 0% |
| Portugal | 0% | 0% | 0% | 0% |
| Romania | 0% | 0% | 10% | 10% |
| Slovakia | 10% | -10% | 10% | 10% |
| Slovenia | 10% | 0% | 10% | 20% |
| Spain | 10% | -10% | 10% | 10% |
| Sweden | 10% | 0% | -10% | 0% |
| United Kingdom | 0% | 10% | -10% | 0% |

*Source: Adapted from Nordregio (2004), Eurostat and WEF, own estimations*

---

[31] The geographic landscape factor is based on the percentage of the land area covered by mountainous municipalities.

[32] The housing socio-demographic factor is based on the distribution of population by dwelling type from Eurostat - *Distribution of population by degree of urbanisation, dwelling type and income group – 2016,* see http://appsso.eurostat.ec.europa.eu/nui/show.do?dataset=ilc_lvho01&lang=en

[33] The national/local regulation factor is based on data from Pillar 1 (Political and regulatory environment) on the Network Readiness Index – 2016, see http://reports.weforum.org/global-information-technology-report-2016/networked-readiness-index/

# References

5G-PPP Automotive Working Group. (2018). A study on 5G V2X Deployment. 5G-PPP Initiative.

Abrardi, L., & Cambini, C. (2019a). Ultra-fast broadband investment and adoption: A survey. *Telecommunications Policy*, *43*(3), 183–198. https://doi.org/10.1016/j.telpol.2019.02.005

Abrardi, L., & Cambini, C. (2019b). Ultra-fast broadband investment and adoption: A survey. *Telecommunications Policy*, *43*(3), 183–198. https://doi.org/10.1016/j.telpol.2019.02.005

Analysis Mason. (2016). Costing the new potential connectivity needs. European Commission. DG Connect. https://doi.org/10.2759/213341

Analysis Mason, & Tech4I2. (2013). *The socio-economic impact of bandwidth*. Brussels.

Analysys Mason. (2016). *Costing the new potential connectivity needs*. https://doi.org/10.2759/213341

Bock, W., & Wilms, M. (2016). Building the Gigabit Society: An Inclusive Path Toward Its Realization. The Boston Consulting Group.

Briglauer, W. (2014). The impact of regulation and competition on the adoption of fiber-based broadband services: Recent evidence from the European union member states. *Journal of Regulatory Economics*, *46*(1). https://doi.org/10.1007/s11149-013-9237-4

Briglauer, W., Stocker, V., & Whalley, J. (2020). Public policy targets in EU broadband markets: The role of technological neutrality. *Telecommunications Policy*, *44*(5), 101908. https://doi.org/10.1016/j.telpol.2019.101908

Cambini, C., & Jiang, Y. (2009). Broadband investment and regulation: A literature review. *Telecommunications Policy*, *33*(10–11), 559–574. https://doi.org/10.1016/j.telpol.2009.08.007

Charalampopoulos, G., Katsianis, D., & Varoutas, D. (2020). Investigating the intertwining impact of wholesale access pricing and the commitment to net neutrality principle on European next-generation access networks private investment plans: An options-game application for capturing market players' competitive in. *Telecommunications Policy*, *44*(3), 101940. https://doi.org/10.1016/j.telpol.2020.101940

Chiha, A., Van der Wee, M., Colle, D., & Verbrugge, S. (2020). Techno-economic viability of integrating satellite communication in 4G networks to bridge the broadband digital divide. *Telecommunications Policy*, *44*(3), 101874. https://doi.org/10.1016/j.telpol.2019.101874

EIB. (2011). *Investigation of the Telecommunications Investment Targets*. Luxembourg.

Enck, J., & Reynolds, T. (2009). Network Developments in Support of Innovation and User Needs. *OECD Digital Economy Papers*, (164), 0_1,2,4-70. Retrieved from http://search.proquest.com/docview/189840148?accountid=10673%5Cnhttp://ope

nurl.ac.uk/athens:_edu?url_ver=Z39.88-2004&rft_val_fmt=info:ofi/fmt:kev:mtx:journal&genre=unknown&sid=ProQ:ProQ%3Aabiglobal&atitle=Network+Developments+in+Support+of+Innovation+and+

European Commission. (2010). A Digital Agenda for Europe. Brussels: COM(2010) 245.

European Commission. (2016). Connectivity for a Competitive Digital Single Market - Towards a European Gigabit Society. Brussels, 14.9.2016: COM(2016) 587 final.

European Court of Auditors. (2018). *Broadband in the EU Member States: despite progress, not all the Europe 2020 targets will be met*. Luxembourg: Publication Office of the European Union. Retrieved from https://www.eca.europa.eu/Lists/ECADocuments/SR18_12/SR_BROADBAND_EN.pdf

FCC. (2010). The broadband availabilty gap. OBI technical paper No 1. Washington: Federal Communications Commission.

Feijóo, C., & Gómez-Barroso, J. L. (2013). El despliegue de redes de acceso ultrarrápidas: un análisis prospectivo de los límites de mercado. *Papeles de Economía Española*, *136*, 116–129.

Feijóo, C., Gómez-Barroso, J. L., & Bohlin, E. (2011, October). Public support for the deployment of next generation access networks - Common themes, methodological caveats and further research. *Telecommunications Policy*. Elsevier Ltd. https://doi.org/10.1016/j.telpol.2011.08.002

Feijóo, C., Ramos, S., Armuña, C., Arenal, A., & Gómez-Barroso, J. L. (2018). A study on the deployment of high-speed broadband networks in NUTS3 regions within the framework of digital agenda for Europe. *Telecommunications Policy*, *42*(9), 682–699. https://doi.org/10.1016/j.telpol.2017.11.001

FTTH Council Europe. (2017). *The Cost of Meeting Europe's Future Network Needs*. Retrieved from http://ftthcouncil.eu/documents/Reports/2017/FTTH Council Cost Model 2017_final.pdf

Galloway, L. (2007). Can broadband access rescue the rural economy? *Journal of Small Business and Enterprise Development*, *14*(4), 641–653. https://doi.org/10.1108/14626000710832749

Given, J. (2010). Take your partners: Public private interplay in Australian and New Zealand plans for next generation broadband. *Telecommunications Policy*, *34*(9), 540–549. https://doi.org/10.1016/j.telpol.2010.07.012

Gomez-Barroso, J. L., & Feijóo, C. (2010). A conceptual framework for public-private interplay in the telecommunications sector. *Telecommunications Policy*, *34*(9), 487–495. https://doi.org/10.1016/j.telpol.2010.01.001

Gómez-Barroso, J. L., & Feijóo, C. (2009). Policy tools for public involvement in the deployment of next generation communications. *Info*, *11*(6), 3–13. https://doi.org/10.1108/14636690910996687

Gómez-Barroso, J. L., & Feijóo, C. (2012). Volition versus feasibility: State aid when aid is looked upon favourably: The broadband example. *European Journal of Law and Economics*, *34*(2), 347–364. https://doi.org/10.1007/s10657-010-9159-x

Gruber, H., Hätönen, J., & Koutroumpis, P. (2014). Broadband access in the EU: An assessment of future economic benefits. *Telecommunications Policy*, *38*(11), 1046–1058. https://doi.org/10.1016/j.telpol.2014.06.007

Han, S., Sung, K. W., & Zander, J. (2013). An Economic Cost Model for Network Deployment and Spectrum in Wireless Networks. In *24th European Regional International Telecommunications Society Conference*. Florence. https://www.econstor.eu/bitstream/10419/88512/1/773874844.pdf

Hätönen, J. (2011). The economic impact of fixed and mobile high-speed networks. *European Investment Bank Papers*, *16*(2), 31–59. Retrieved from http://hdl.handle.net/10419/54666

Hernández, P. F., Aguilar Chinea, R. M., & Baquero Pérez, P. (2020). Public aid for ultra-fast broadband development in archipelagos. The Canary Islands case. *Digital Policy, Regulation and Governance*, *22*(3), 201–225. https://doi.org/10.1108/DPRG-12-2019-0113

Hong, J. pyo. (2017). Causal relationship between ICT R&D investment and economic growth in Korea. *Technological Forecasting and Social Change*, *116*. https://doi.org/10.1016/j.techfore.2016.11.005

IDATE. (2013). *Deployment costs & access market revenue*. Montpellier.

IHS Markit, Point Topic, & Omdia. (2020). *Broadband Coverage in Europe 2019. Mapping progress towards the coverage objectives of the Digital Agenda : final report*. https://doi.org/https://doi.org/10.2759/375483

IHS Markit, & Point Topic. (2018). *Broadband coverage in Europe 2017: final report*. https://doi.org/https://doi.org/10.2759/358688

Ioannou, N., Katsianis, D., & Varoutas, D. (2020). Comparative techno-economic evaluation of LTE fixed wireless access, FTTdp G.fast and FTTC VDSL network deployment for providing 30 Mbps broadband services in rural areas. *Telecommunications Policy*, *44*(3), 101875. https://doi.org/10.1016/j.telpol.2019.101875

Lebourges, M., & Liang, J. (2018). Estimating the impact of co-investment in fiber to the home coverage. *Econstor*.

Lemstra, W., & Melody, W. H. (2015). Cross-case analysis. In W. Lemstra, W. Melody, M. Van der Wee, & S. Verbrugge (Eds.), *The dynamics of broadband markets in Europe. Realizing the 2020 Digital Agenda* (p. 409). Cambridge University Press. https://www.researchgate.net/publication/273760658_The_Dynamics_of_Broadband_Markets_in_Europe_-_Realizing_the_2020_Digital_AGenda

Logothetis, V., Ioannou, N., Tselekounis, M., Katsianis, D., Chipouras, A., & Varoutas, D. (2019). Ex post regulation of NGA networks: A roadmap for calculating the cost of an efficient operator – The case of Greece. *Econstor*.

Lucendo-Monedero, A. L., Ruiz-Rodríguez, F., & González-Relaño, R. (2019). Measuring the digital divide at regional level. A spatial analysis of the inequalities in digital development of households and individuals in Europe. *Telematics and Informatics*, *41*, 197–217. https://doi.org/10.1016/j.tele.2019.05.002

Oughton, E. J., & Frias, Z. (2016). The cost, coverage and rollout implications of 5G infrastructure in Britain. *ITRC Mistral*.

Ovando, C., Pérez, J., & Moral, A. (2015). LTE techno-economic assessment: The case of rural areas in Spain. *Telecommunications Policy*, *39*(3), 269–283. https://doi.org/10.1016/j.telpol.2014.11.004

Point Topic. (2013). *Europe's broadband investment needs: Quantifying the investment needed to deliver superfast broadband to Europe*. London.

Queder, F. (2020a). Competitive effects of cable networks on FTTx deployment in Europe. *Telecommunications Policy*, *44*(10). https://doi.org/10.1016/j.telpol.2020.102027

Queder, F. (2020b). Towards a vertically separated broadband infrastructure: The potential role of voluntary separation. *Competition and Regulation in Network Industries*, *21*(2), 143–165. https://doi.org/10.1177/1783591720907004

Rendon Schneir, J., & Xiong, Y. (2016). A cost study of fixed broadband access networks for rural areas. *Telecommunications Policy*, *40*(8), 755–773. https://doi.org/10.1016/j.telpol.2016.04.002

Ruhle, E. O., Brusic, I., Kittl, J., & Ehrler, M. (2011). Next Generation Access (NGA) supply side interventions - An international comparison. *Telecommunications Policy*, *35*(9–10), 794–803. https://doi.org/10.1016/j.telpol.2011.06.001

Sahebali, M. W. W., Sadowski, B. M., Nomaler, O., & Brennenraedts, R. (2019). Infrastructure Rollout and Fibre Provision: The case of NGN in the Netherlands. *Econstor*.

Salemink, K., & Strijker, D. (2018). The participation society and its inability to correct the failure of market players to deliver adequate service levels in rural areas. *Telecommunications Policy*, *42*(9), 757–765. https://doi.org/10.1016/j.telpol.2018.03.013

Taga, K., & Peres, G. (2018). The race to gigabit fiber. Gigabit fiber is driving take-up of new services. Arthur D. Little.

Tselekounis, M., & Maniadakis, D. (2012). NGA Investments: A departure from the existing cost and demand structure assumptions. In International Telecommunications Society (Ed.), *Proceedings of the 19th ITS Biennial Conference*. Bangkok, Thailand.

Vodafone. (2017). *Vodafone Germany GIGABIT INVESTMENT PLAN*. https://www.vodafone.com/news/press-release/vodafone-germany-gigabit-investment-plan